\documentclass[twocolumn,prb,amsmath,amssymb,longbibliography,floatfix,superscriptaddress]{revtex4-2}
\usepackage{amsmath}
\usepackage{amssymb}
\usepackage{color}
\usepackage{physics}
\usepackage{graphicx}
\usepackage{tabularx}
\usepackage{dcolumn} 
\usepackage{bm}      
\usepackage{array}   
\usepackage{amsfonts}
\usepackage{url}
\usepackage[bookmarks=false,colorlinks,citecolor=red]{hyperref}

\usepackage[color=green!60]{todonotes}
\usepackage[utf8]{inputenc}
\usepackage[T1]{fontenc}

\include{MyCommand}

\newcommand{\avg}[1]{\expval{#1}}

\DeclareUnicodeCharacter{2212}{-}

\newcommand{\aidag}[1]{\hat{a}_{#1}^{\dag}}
\newcommand{\ai}[1]{\hat{a}_{#1}^{\phantom\dag}}

\newcommand{\effspec}[1]{\tilde{\epsilon}_{#1}}
\newcommand{\overlap}[2]{\langle #1 | #2 \rangle}
\newcommand{\obs}[1]{\langle #1 \rangle}

\usepackage{xr} 

\makeatletter
\newcommand*{\addFileDependency}[1]{
\typeout{(#1)}
\@addtofilelist{#1}
\IfFileExists{#1}{}{\typeout{No file #1.}}
}\makeatother

\newcommand*{\myexternaldocument}[1]{%
\externaldocument{#1}%
\addFileDependency{#1.tex}%
\addFileDependency{#1.aux}%
}

\myexternaldocument{long_supplement}

\usepackage{pdfpages} 
\usepackage{pgffor} 

\makeatletter
\AtBeginDocument{\let\LS@rot\@undefined}
\makeatother

\def\supplementfilename{long_supplement}

\pdfximage{\supplementfilename.pdf}
\def\numbersupplementpages{\the\pdflastximagepages}

\newif\ifarXiv
\arXivtrue 

\begin{document}

\preprint{AIP/123-QED}

\title{A Stable, Recursive Auxiliary Field Quantum Monte Carlo Algorithm in the Canonical Ensemble: Applications to Thermometry and the Hubbard Model}

\author{Tong Shen}
 \affiliation{Department of Chemistry, Brown University, Providence, RI 02912
 }
  \author{Hatem Barghathi}
 \affiliation{Department of Physics and Astronomy, University of Tennessee, Knoxville, TN 37916
 }
 \author{Jiangyong Yu}
 \affiliation{Department of Physics, Brown University, Providence, RI 02912
 }
 \author{Adrian Del Maestro}
 \affiliation{Department of Physics and Astronomy, University of Tennessee, Knoxville, TN 37916
 }
 \affiliation{Min H.~Kao Department of Electrical Engineering and Computer Science, University of Tennessee, Knoxville, TN 37996, USA}

\author{Brenda M. Rubenstein}
\email{Author to whom correspondence should be addressed: brenda\_rubenstein@brown.edu.}
 \affiliation{Department of Chemistry, Brown University, Providence, RI 02912
 }
  \affiliation{Department of Physics, Brown University, Providence, RI 02912
 }

\date{\today}
\begin{abstract}
Many experimentally-accessible, finite-sized interacting quantum systems are most appropriately described by the canonical ensemble of statistical mechanics. Conventional numerical simulation methods either approximate them as being coupled to a particle bath, or use projective algorithms which may suffer from non-optimal scaling with system size or large algorithmic prefactors.  In this paper, we introduce a highly stable, recursive Auxiliary Field Quantum Monte Carlo approach that can directly simulate systems in the canonical ensemble.  We apply the method to the fermion Hubbard model in one and two spatial dimensions in a regime known to exhibit a significant ``sign'' problem and find improved performance over existing approaches including rapid convergence to ground state expectation values.  The effects of excitations above the ground state are quantified using an estimator-agnostic approach including studying the temperature dependence of the purity and overlap fidelity of the canonical and grand canonical density matrices.  As an important application, we show that thermometry approaches often exploited in ultra-cold atoms that employ an analysis of the velocity distribution in the grand canonical ensemble may be subject to errors leading to an under-estimation of extracted temperatures with respect to the Fermi temperature.
\end{abstract}

\keywords{Canonical Ensemble, Auxiliary Field Quantum Monte Carlo, Determinant Quantum Monte Carlo, Auxiliary Partition Functions, Finite Temperature, Thermometry, Hubbard Model}

\maketitle

\section{Introduction}
In many settings in condensed matter physics, the grand canonical ensemble, in which the number of particles in a system is allowed to fluctuate subject to a fixed chemical potential, is the ensemble of choice for modeling systems at finite temperature. This is a natural framework due to the approach to the thermodynamic limit for electrons in solids, or the existence of a particle (or quasi-particle) reservoir in transport geometries, heterostructures, and superconductors. However, there are a growing number of important scenarios in which the number of particles is fixed and small, including trapped atom systems comprised of a finite number of atoms \cite{Mukherjee:2017mz,Hueck:2018td}, nuclear systems with a fixed number of nucleons \cite{Alhassid_Book}, and molecules containing a fixed number of electrons \cite{Liu_JCTC_2018,Petras_JCTC}. All such systems are more accurately described by the canonical ensemble in which the number of particles cannot fluctuate. Moreover, many ground state algorithms in condensed matter are formulated in the canonical ensemble \cite{Malone_JCP_2015,Malone_PRL_2016,Brown_PRL_2013,Zhang_PRL_2003} and finite temperature algorithms that can converge to these algorithms' ground state results without spurious particle number fluctuations can shed a brighter light on the mechanisms behind low-temperature quantum phase transitions and crossovers \cite{Liu_JCTC_2020,Petras_JCTC}. Other examples where systems need to be treated within the canonical ensemble include determining the operationally-accessible entanglement in indistinguishable many-body systems in the presence of a $U(1)$ superselection rule limiting physically allowable operations \cite{Wiseman:2003ei,Bartlett:2003ec,Barghathi:2018rg} and the determination of thermonuclear rates for astrophysics \cite{Rauscher:1997ld}.

Efficiently describing interacting systems in the canonical ensemble has nevertheless been a longstanding challenge, particularly for second-quantized algorithms. Unlike in the grand canonical ensemble, in which partition functions and other quantities can be evaluated without placing any constraints on the number of particles \cite{mcquarrie2000statistical}, evaluating quantities in the canonical ensemble requires an explicit consideration of particle-number constraints. At a physical level, these constraints give rise to interesting, nontrivial correlations among the occupations of different states -- higher order expectation values of occupation numbers do not factorize, even in the non-interacting limit \cite{barghathi2020theory}. However, at a numerical level, these constraints can make the analytical and computational evaluation of canonical ensemble quantities substantially more cumbersome \cite{borrmann1993recursion}.  

One approach for modeling interacting systems in the canonical ensemble that circumvents the imposition of direct constraints is the use of projection techniques \cite{Ormand_PRC,Gilbreth_CompPhysComm_2015}. In these algorithms, canonical ensemble quantities are projected out from the grand canonical partition function at a suitably tuned chemical potential \cite{Miles:2022hr}. This approach has been fruitfully employed to study a wide variety of problems in nuclear physics \cite{Lang_PRC_1993, Ormand_PRC}, and more recently, condensates \cite{Gilbreth_PRA,Jensen_PRL_2020}. Nonetheless, because this algorithm relies upon projecting out of the grand canonical ensemble, it is accompanied by the same computational overhead as typical grand canonical simulations and can develop numerical instabilities for large particle numbers or when reasonable chemical potentials cannot be identified, for example near first order phase transitions \cite{Herdman:2014cp}. Recently, techniques for rapidly determining a chemical potential where it can be readily identified have been proposed, but these techniques still do not inherently operate in the canonical ensemble \cite{Wang_Toldin_PRE_2017,Miles:2022hr,Sedgewick_PRB_2003}. Approaches that directly take physical constraints into account therefore have the potential to lead to methods that are not only more stable, but also more computationally efficient. 

Recently, a new Auxiliary Partition Function (APF) formalism has been proposed that enables the recursive computation of $N$-particle partition functions and related quantities for non-interacting systems from smaller particle number quantities, thus explicitly taking particle-number constraints into account \cite{barghathi2020theory}. In essence, the APF formalism views the canonical ensemble partition function as a sum over the probabilities of varying numbers of particles occupying different subsets of states. Unlike previous such recursions for non-interacting systems \cite{borrmann1993recursion}, the APF formalism is able to arrive at expressions for canonical ensemble partition functions using only positive sub-quantities, and hence it is significantly more numerically stable. While this technique has been successfully applied to such non-interacting systems as harmonic oscillators and rings of bosons \cite{barghathi2020theory}, it and alternative recursions have yet to be generalized to interacting systems. 

In this manuscript, we present a new, highly stable recursive algorithm for the simulation of interacting systems in the canonical ensemble. This algorithm marries the finite temperature Auxiliary Field Quantum Monte Carlo (AFQMC) (or Determinant Quantum Monte Carlo (DQMC)) algorithm \cite{White_PRB_1989,Zhang_PRL_1999}, which has long been used to model finite temperature interacting systems, with the Auxiliary Partition Function (APF) formalism \cite{barghathi2020theory}. The key realization that enables this marriage is the fact that, through the Hubbard-Stratonovich (HS) Transformation \cite{Hirsch_PRB_1983}, the AFQMC algorithm reconstructs the properties of interacting systems by integrating over an appropriately-weighted ensemble of non-interacting systems \cite{Motta_WIRES}. Our algorithm can therefore stably describe interacting systems in the canonical ensemble by sampling non-interacting partition functions and other finite temperature quantities generated using the APF formalism and then integrating over those samples. This markedly improves upon our previous work which was built upon the significantly less stable Borrmann recursion algorithm for non-interacting gases \cite{borrmann1993recursion,Shen_JCP_2020}. To highlight the stability of our new algorithm, we show that (1) our new interacting algorithm is stable down to substantially lower temperatures than previous algorithms and (2) it has a lower computational scaling than previous Projection algorithms. 

With this highly stable algorithm, we proceed to analyze differences in the convergence of the energy, sign, and information theoretic measures such as the purity and fidelity \cite{Bonnes_PRB_2013,Gilles2022_PRB} to the ground state between the grand canonical and canonical ensembles. To do so, we focus on the Hubbard model of interacting fermions in one and two dimensions as an instructive example because it manifests the strong correlation often resulting in a sign problem that is hardest to model via modern simulation techniques. We find that because higher-energy states are more readily accessed in the grand canonical ensemble, grand canonical energies tend to be higher and purities lower at any given temperature, meaning that the grand canonical ensemble converges more slowly to the ground state. We substantiate these findings with analytical expressions describing how these quantities should converge to the ground state in both the interacting and non-interacting limits. We furthermore demonstrate that these differences have substantial practical implications for the thermometry of cold atom systems: if the temperature of cold atom systems containing a fixed number of particles is estimated based on the grand canonical ensemble, this leads to temperature predictions that can be up 53.2\% lower than in the more realistic canonical ensemble picture according to the analysis we present below.  

We begin in Section \ref{formalism} by presenting our new algorithm and its underlying formalism, showing how recursive relations for the partition function and one- and two-body quantities can be determined using the APF method and subsequently integrated into the AFQMC algorithm. We also demonstrate the formal relationship between our recursive algorithm and the previously used Projection algorithm. In Section \ref{results}, we present our results regarding the increased stability of our algorithm, before illustrating the differences in system energies, purities, and fidelities as measured in the two ensembles using the Hubbard model as a salient example. We exemplify the practical consequences of these differences for thermometry in Section \ref{thermometry}. We lastly conclude in Section \ref{conclusions} by discussing further applications of our algorithm and its straightforward extension to studying nuclear matter and bosons, for which it has the potential to show even greater efficiency gains.

\section{Formalism \label{formalism}}

\subsection{\label{sec:afqmc} The Finite Temperature Auxiliary Field Quantum Monte Carlo (AFQMC) Algorithm} 
 
The central quantity in finite temperature theories is the partition function, from which all other properties can be derived. Historically, the focus of finite temperature methods has been to obtain or otherwise sample from the grand partition function, $\mathcal{Z}_{\mu}$, associated with the grand canonical  ensemble, in which the internal energy and particle number are allowed to fluctuate around average values that can be tuned by the temperature $T$ and chemical potential $\mu$, respectively \cite{mcquarrie2000statistical}. 

The grand partition function can be expressed as the trace
\begin{equation}
    \mathcal{Z}_{\mu} = \text{Tr} \left(e^{-\beta (\hat{H} - \mu \hat{N})} \right)\, 
    \label{eqn:GrandPartitionFunction}
\end{equation}
where $\beta=1/(k_{\rm B}T)$, $k_{\rm B}$ denotes the Boltzmann constant, $\hat{H}$ denotes the many-body Hamiltonian, and $\hat{N}$ is an operator corresponding to the total number of particles which is the sum over the occupations over the set of (possibly degenerate) states.  In order to facilitate its subsequent sampling, the grand partition function can be discretized into $L$ imaginary time slices, each of which can then be approximately factored into short imaginary time one- and two-body propagators via a Suzuki-Trotter factorization \cite{Trotter,Suzuki_Prog1976}
\begin{eqnarray}
\mathcal{Z}_{\mu} &=& \text{Tr} \left(\prod_{l=1}^L e^{-\Delta \tau (\hat{H} - \mu \hat{N})}\right) \nonumber \\
&\approx& \text{Tr} \left( \prod_{l=1}^L [e^{-\Delta \tau \hat{K} / 2} e^{-\Delta \tau \hat{V} } e^{-\Delta \tau \hat{K} / 2}] \right),  
\end{eqnarray}
where $\Delta \tau = \beta/L$ represents an imaginary time slice, $\hat{K}$ is the collection of all one-body operators, and $\hat{V}$ is the collection of all two-body operators. The exact grand partition function is recovered in the limit $\Delta \tau \rightarrow 0$. This factorization enables us to ignore the non-commutation of the one-body and two-body operators up to a systematic and controllable error of $O(\Delta \tau^3)$. 

While one-body propagators, $e^{-\Delta \tau \hat{K}}$, may be neatly expressed as matrices in a given basis \cite{Hirsch_PRB}, two-body propagators, $e^{-\Delta \tau \hat{V}}$, may not be as easily determined. In auxiliary field-based methods, including Determinant Quantum Monte Carlo (DQMC) \cite{White_PRB_1989,Bai_Chapter} and Auxiliary Field Quantum Monte Carlo (AFQMC) \cite{Zhang_PRL_2003,rubenstein2012finite}, two-body propagators of the form $e^{- \Delta \tau \hat{V}}$ are linearized by re-expressing them as integrals over one-body propagators \cite{Hirsch_PRB_1983}
using the Hubbard-Stratonovich Transformation \cite{Hirsch_PRB_1983,Buendia_PRB}. Assuming as we will below that the two-body propagator can be written as 
\begin{equation}
    \hat{V} = \sum_{\gamma} \lambda_{\gamma} \hat{v}_{\gamma}^2,
\end{equation}
where the $\hat{v}_{\gamma}$ denote linear combinations of one-body operators and the $\lambda_{\gamma}$ denote their contributions to the sum, then 
\begin{equation}
    e^{-\Delta \tau \hat{V}} = \int D\sigma G_{\sigma} \hat{V}_{\sigma}
    \label{HSTransform}
\end{equation}
according to the HS Transformation, where $\sigma$ represents an auxiliary field, $G_{\sigma}$ represents the Gaussian probability of sampling that field, and $\hat{V}_{\sigma}$ denotes the collection of one-body operators as a function of that field. Crucial to this paper, Eq.~\eqref{HSTransform} signifies that propagators for interacting systems can be rewritten as integrals over propagators for non-interacting systems. Based on this transform, all of the one-body operators from different time slices, $i$, are then combined, i.e., $\hat{U}_{\sigma} = \prod_i e^{-\Delta \tau \hat{K} / 2} \hat{V}_{\sigma_{i}} e^{-\Delta \tau \hat{K} / 2}$, and the full grand canonical partition function may be expressed as 
\begin{eqnarray} 
\mathcal{Z}_{\mu} &=& \text{Tr} \left(\int D\sigma G_{\sigma} e^{\beta \mu \hat{N}} \hat{U}_{\sigma} \right) \nonumber \\
&=&  \int D\sigma G_{\sigma} \text{Tr} \left( e^{\beta \mu \hat{N}} \hat{U}_{\sigma} \right) \nonumber \\
&=& \int D\sigma G_{\sigma} \text{det}(I+e^{\beta \mu} \textbf{U}_{\sigma} ), \label{Eqn:PartitionFuncGCE}
\end{eqnarray} 
where taking the trace over all fermion occupations results in a determinant \cite{Hirsch_PRB}. 

The partition function can then be sampled to evaluate such observables as energies, average occupations, and correlation functions \cite{White_PRB_1989,Zhang_PRL_1999}. In particular, Wick's Theorem is valid in the grand canonical ensemble, which enables a powerful simplification of expectation values of products of operators (e.g., correlation functions) into factorized sums and differences of shorter products of expectation values of those operators \cite{fetter2012quantum,Wick:1950wt}.

\subsection{Recursive Relations for the Canonical Partition Function} 

While computing properties in the grand canonical ensemble is appropriate for many systems and can be analytically/computationally convenient, in many situations in which the particle number remains fixed, a canonical treatment is more suitable. Computing the canonical ensemble partition function proceeds along the same lines as computing the grand canonical one with the critical exception that the trace must be taken with the constraint of fixed particle number $N$. More specifically, the $N$-particle, canonical ensemble partition function may be expressed as
\begin{equation}
    Z_N = \text{Tr}_N (e^{-\beta \hat{H}}),
\end{equation}
which can be factored and transformed in a similar manner to the grand partition function to obtain
\begin{equation}
    \label{FirstZn}
    Z_N = \int D\sigma G_{\sigma} \text{Tr}_N (\hat{U}_{\sigma}), 
\end{equation}
where we have added the subscript $N$ to the trace to differentiate it from that in Eq.~\eqref{eqn:GrandPartitionFunction}. Because $\hat{U}_{\sigma}$ is a one-body operator, its matrix form, $\bf U_{\sigma}$, can be diagonalized in the single-particle space:
\begin{equation}
    \bf U_{\sigma} = P \Lambda P^{-1},
    \label{eqn:AExpansion}
\end{equation}
where we omit the $\sigma$-depedence on the right side for clarity. We then introduce the effective single-particle spectrum $\Lambda = {\rm diag}(\{\lambda_\gamma\}) = {\rm diag}(\{\exp(-\beta \effspec{\gamma}) \})$, based upon the following relations
\begin{equation}
    \hat{U}_{\sigma} = \sum_{\gamma} \exp(-\beta \aidag{\gamma} \effspec{\gamma} \ai{\gamma}) = \sum_{\gamma} \lambda_{\gamma}^{\hat{n}_{\gamma}},
\end{equation}
and the basis transformation
\begin{equation}
    \aidag{\gamma} = \sum_{i} \overlap{i}{\gamma} \aidag{i}, \hspace{0.3 cm} \ai{\gamma} = \sum_{i} \overlap{\gamma}{i} \ai{i}, \hspace{0.3 cm} \hat{n_\gamma} = \aidag{\gamma} \ai{\gamma}.
    \label{eqn:BasisRotation}
\end{equation}
Since $\hat{U}_{\sigma}$ is an independent-particle propagator that only depends on the auxiliary field vector, $\sigma$, the effective single-particle spectrum, $\{ \effspec{\gamma} \}$, is independent of the particle number. For an $N$-particle, $N_s$-site system, taking the trace while constraining the particle number yields \cite{Shen_JCP_2020}
\begin{align}
    \text{Tr}_N (\hat{U}_\sigma) &= \text{Tr}_N (\sum_{\gamma} \exp(-\beta \aidag{\gamma} \effspec{\gamma} \ai{\gamma})) \nonumber \\
    &= \sum_{\Gamma_N} \langle \Gamma_N | \sum_{\gamma} \exp(- \beta \aidag{\gamma} \effspec{\gamma} \ai{\gamma}) | \Gamma_N \rangle \nonumber \\
    &= \sum_{\Gamma_N} \sum_{\gamma} \lambda_{\gamma}^{n_{\gamma}}. \label{eqn:CanonicalTrace3}
\end{align}
Here, $\Gamma_N$ is used to represent the set of $N$-particle states, and thus, $\sum_{\Gamma_N} \equiv \sum_{n_1 + \dots + n_{N_s} = N}$ and $n_\gamma$ denotes the number of particles in the $\gamma^{\rm th}$ eigenstate. For fermions, $n_\gamma = 0, 1$. The key implication of Eq.~\eqref{eqn:CanonicalTrace3} is that, for a specified field $\sigma$, the single-particle spectrum can be decoupled from the particle number. Hence, the many-particle energy given such fields is simply the sum of all of the single-particle energies. 

This key fact enables us to move beyond previous projection-based approaches and calculate Eq.~\eqref{eqn:CanonicalTrace3} in a recursive fashion, where we utilize the recursive approach to calculating canonical ensemble partition functions first developed for ideal gases. Specifically, the  partition function can be obtained using the well-known Borrmann recursion \cite{borrmann1993recursion,borrmann1999calculation}
\begin{equation}
    Z_N = \sum_{k = 1}^N (\zeta)^{k - 1} z_k Z_{N-k},
\label{eqn:borrmann}
\end{equation}
where $z_k=\sum_{j=1}^{N_s} \lambda_j^{k}$ and $\zeta$ takes the values of $-1$ and $1$ for fermions and bosons, respectively. However, in the context of non-interacting gases, the fermionic version of Eq.~\eqref{eqn:borrmann} is known to suffer from numerical instabilities \cite{schonhammer2017deviations,Schmidt_PA}, which weree also encountered in a previous version of our canonical AFQMC algorithm  \cite{Shen_JCP_2020}, leading to the emergence of an additional \emph{unphysical} sign problem.

We can gain some intuitive understanding of the reasons behind such numerical instability for the case of fermions at low temperatures. In addition to the alternating signs in Eq.~\eqref{eqn:borrmann}, the contribution of high energy levels to the factors $z_k$, can be filtered out by limited numerical precision. This is expected to have a minimal effect on low-temperature Bose gases, as their thermodynamic properties rely heavily on the occupation of low-energy levels. In contrast, the thermodynamic properties of low-temperature Fermi gases are governed by much higher energy levels in the vicinity of the Fermi level. 

Here, we build upon our previous work \cite{Shen_JCP_2020} to propose a more numerically stable and accurate method for computing the canonical trace via a recursive formula based upon the recently developed Auxiliary Partition Functions formalism \cite{barghathi2020theory}. Given a set $\{\lambda_i=e^{-\beta\effspec{i}}\}$ of Boltzmann factors that correspond to the single particle energy spectrum $\{\effspec{i}\}$, we can build the desired $N$ particle partition function recursively by including one of the levels in each recursive step, i.e.,
\begin{equation}
    Z_{N} = \lambda_j Z_{N-1}^{\{\lambda_i\} \backslash \lambda_j} + Z_{N}^{\{\lambda_i\} \backslash \lambda_j}\ , \label{Eqn:RecursionForZ}
\end{equation}
where the notation $\qty{\lambda_i} \backslash \lambda_j$ implies that we exclude the specific level $j$ from the set $\qty{\lambda_i}$ (see Ref.~\cite{barghathi2020theory} for extensive details).  This recursion does not suffer from the problems of  Eq.~\eqref{eqn:borrmann} because it is inherently positive. Yet, the unbounded nature of the Boltzman factors $\lambda_i$ can make achieving the desired numerical precision difficult, especially at low temperatures. This can be addressed with a simple trick. We modify Eq.~\eqref{Eqn:RecursionForZ} by inserting an arbitrary multiplicative factor $A_j$ (to be determined below) with the inclusion of each $\lambda_j$ and apply the modified equation on the modified set $\{B\lambda_i\}$, where $B$ is an additional constant (also to be determined).  This results in a modified recursion relation
\begin{equation}
    \bar{Z}_{N} = A_j\left(B\lambda_j\bar{Z}_{N-1}^{\{\lambda_j\} \backslash \lambda_j} + \bar{Z}_{N}^{\{\lambda_j\} \backslash \lambda_j}\right), \label{Eqn:RecursionForZmod}
\end{equation}
and we can recover $Z_{N}$ from the resulting $\bar{Z}_{N}$ via 
\begin{equation}
    \bar{Z}_{N} =B^NZ_{N}\prod_jA_j. \label{Eqn:ZandZmod}
\end{equation}
This suggests that we can enhance the performance of the APF recursive approach for calculating $Z_{N}$ through a clever choice of the constants $\{A_j\}$ and $B$. 
In order to do so, we rearrange the fugacity expansion for the grand canonical partition function, $\mathcal{Z}_{\mu}$, in terms of the canonical partition functions, $Z_{N} = \Tr_N e^{-\beta \hat{H}}$,
\begin{equation}
    \mathcal{Z}_{\mu} = \sum_N e^{\beta \mu N} Z_{N}.
\end{equation}
Dividing by the grand canonical partition function and removing the summation over $N$ results in an expression for the particle number probability distribution given by 
\begin{equation}
 P_{\mu}(N) = \frac{e^{\beta \mu N} Z_{N}}{\mathcal{Z}_{\mu}}\ . \label{Def:PN}
\end{equation}
For non-interacting fermions,  $\mathcal{Z}_{\mu}=\prod_j\left(1-p_j^{(\mu)}\right)^{-1}$ \cite{mcquarrie2000statistical}, where
\begin{equation}
    p_j^{(\mu)} = \frac{e^{\beta \mu}\lambda_j}{1 + e^{\beta \mu}\lambda_j} \label{Def:OccProb}
\end{equation}
is the probability of occupying the $j^{\rm th}$ energy level. This yields 
\begin{equation}
   P_{\mu}(N) = e^{\beta \mu N} Z_{N}\prod_j\left(1-p_j^{(\mu)}\right). \label{Def:PNFreeF}
\end{equation}
If we compare Eq.~\eqref{Def:PNFreeF} with Eq.~\eqref{Eqn:ZandZmod}, we can identify $P_{\mu}(N)$ with $\bar{Z}_{N}$ by choosing $A_j=1-p_j^{(\mu)}$  and $B={e}^{\beta\mu}$, which, when substituted into Eq.~\eqref{Eqn:RecursionForZmod}, results in a recursion relation for the number probability distribution
\begin{equation}
    P_{\mu}(N) = p_j^{(\mu)} P^{\{\lambda_i\} \backslash \lambda_j}_{\mu}(N - 1) + (1 - p_j^{(\mu)}) P^{\{\lambda_i\} \backslash \lambda_j}_{\mu}(N). \label{Eqn:RecursionForPN}
\end{equation}
In contrast with Eq.~\eqref{Eqn:RecursionForZ}, all terms in the above equation
are bounded between $0$ and $1$, which further ensures their numerical
stability and automatically avoids numerical arithmetic overflow issues
caused by extremely large $\lambda_i$ values. Also, setting $N=0$ in Eq.~\eqref{Def:PN}, we see that $1/\mathcal{Z}_{\mu}=P_{\mu}(0)$, which enables us to re-express Eq.~\eqref{Def:PN} as 
\begin{equation}
    Z_N = \frac{Z_N}{Z_0} = \frac{e^{\beta \mu N} P_{\mu}(N)}{P_{\mu}(0)}, \label{Eqn:RecursionForZn}
\end{equation}
where $Z_0 = 1$.

We note that in Eq.~\eqref{Eqn:RecursionForZn}, the chemical potential $\mu$ is an algorithmic parameter that can take on any value without changing the value of the canonical trace. It need not be the many-body chemical potential, which is otherwise difficult to determine for an arbitrary system. To increase the numerical stability of Eqs.~\eqref{Eqn:RecursionForPN} and \eqref{Eqn:RecursionForZn}, it is best to select a $\mu$ around the Fermi level, where $P_{\mu}$ peaks at $N$. A good choice is $e^{\beta \mu} = |\lambda_{N} \lambda_{N+1}|^{1/2}$, assuming that $\{\lambda_i\}$ is sorted as $|\lambda_1| < |\lambda_2| < \cdots < |\lambda_{N_s}|$.

It is worth mentioning that the particle number distribution $\mathcal{P}_{\mu}(N)$ can be viewed as a Poisson-binomial distribution \cite{HONG_2013}, which can be expressed as
\begin{equation}
    P_{\mu}(N) := \sum_{\mathcal{S}_N} \prod_{i \in \mathcal{S}_N} p_i^{(\mu)} \prod_{j \in \bar{\mathcal{S}}_N}(1 - p_j^{(\mu)}), \label{Def:PoissBino}
\end{equation}
where it is constructed from the probability $p_i^{(\mu)}$ of successfully occupying $N$ energy levels at a given chemical potential and temperature out of a total number of independent (non-interacting) and nonidentical Bernoulli trials \cite{barghathi2020theory}, where the corresponding set of independent success probabilities is represented by $\qty{p_i^{(\mu)}}$. Here, $\mathcal{S}_N$ is a set of $N$ occupied energy levels selected from $N_s$ energy levels in the single-particle space; $\bar{\mathcal{S}}_N$ denotes the complement coming from unoccupied levels.

While we have focused our presentation of our formalism here on fermions, it can readily be extended to bosons, as further detailed in the Supplementary Materials \cite{supplemental}. 

\subsection{Relationship between Auxiliary Partition Function Approach and Previous Projection Approaches} 

In this section, we illustrate how the more conventional particle number projection
formalism \cite{Ormand_PRC_1994} can be derived from Eqs.~\eqref{Def:OccProb}, \eqref{Def:PNFreeF}, and \eqref{Eqn:RecursionForPN} through a Fourier transform, and hence that it is analytically equivalent to the recursive approach. The fact that these approaches are analytically equivalent will be of value when we compare the accuracy, stability, and speed of these methods in Section \ref{accuracy}. 

We begin by considering the generating function of the partition function under the recursive relation, Eq. \eqref{Eqn:RecursionForZ}, for any non-zero $a$
\begin{eqnarray}
    \sum_{N=0}^{\infty} a^N Z_{N} &=& \lambda_j \sum_{N=1}^{\infty} a^N Z_{N-1}^{\{\lambda_j\} \backslash \lambda_j} + \sum_{N=0}^{\infty} a^N Z_{N}^{\{\lambda_j \} \backslash \lambda_j}\nonumber \\
    &=& (1 + a \lambda_j) \sum_{N=0}^{\infty} a^N Z_{N}^{\{\lambda_j \} \backslash \lambda_j}.
\end{eqnarray}
By iteratively subtracting the Boltzmann factors $\lambda_j$ from the set of all levels $\{\lambda_i\}$, we arrive at
\begin{equation}
    \sum_{N=0}^{\infty} a^N Z_{N} = \prod_j (1 + a \lambda_j)\, .
\end{equation}
Setting $a = e^{\beta \mu}e^{i \phi_m}$, with $\phi_m = 2 \pi m / N_s$, then yields
\begin{equation}
    \sum_{N=0}^{\infty} e^{\beta \mu N}e^{i \phi_m N} Z_{N} = \prod_j (1 + e^{\beta \mu}e^{i \phi_m} \lambda_j)\, .
\end{equation}
Using the discrete Fourier transform of the delta function at $N$, $\delta_{N} = \sum_{m=1}^{N_s} e^{i \phi_m N} / N_s$, we can solve for $e^{\beta \mu N}Z_{N}$ and recover the particle number projection
result for the canonical trace that was first proposed in Ref.~[\onlinecite{Ormand_PRC_1994}]
\begin{align}
    Z_N &= \frac{e^{-\beta \mu N}}{N_s} \sum_{m = 1}^{N_s} e^{-i \phi_m N} \prod_j (1 + e^{\beta \mu}e^{i \phi_m} \lambda_j) \nonumber \\
    &= \frac{1}{N_s} \sum_{m = 1}^{N_s} e^{-\beta \mu N} e^{-i \phi_m N} \tilde{\mathcal{Z}}_{\mu}(m). \label{Eqn:ProjectionforZN} 
\end{align}

As we shall demonstrate below, the recursion formalism and the projection formalism have equivalent accuracy with
recursion having slightly improved scaling [$O(N_s^3 + N_sN)$ vs.\@ $O(N_s^3 +
N_s^2)$ after considering the $N_s^3$ cost of the eigendecomposition] for computing the
partition function.

\subsection{Recursive Computation of Density Matrices and Correlation Functions} 
The expectation value of the one-body density operator can be evaluated using the same eigendecomposition of the field-dependent propagator matrix, $\textbf{U}_{\sigma} = \textbf{P} \Lambda \textbf{P}^{-1}$, with $\Lambda = \mathrm{diag}(\{ \lambda_i \})$, which leads to
\begin{equation}
    \langle \hat{c}^{\dag}_i \hat{c}_j \rangle_{N} = \sum_{\alpha} \textbf{P}_{i \alpha} \langle \hat{n}_{\alpha} \rangle_{N} \textbf{P}^{-1}_{\alpha j},
\end{equation}
where $\langle \hat{n}_{\alpha} \rangle_{N}$ is computed recursively in $O(N)$ operations \cite{schmidt1989simple,borrmann1999calculation} as
\begin{equation}
    \langle \hat{n}_{\alpha} \rangle_{N} = \frac{\lambda_{\alpha} P_{\mu}(N - 1)}{e^{\beta \mu}P_{\mu}(N)}(1 - \langle \hat{n}_{\alpha} \rangle_{N - 1}) \label{Eqn:RecursionForN}
\end{equation}
with the initial condition $\langle \hat{n}_{\alpha} \rangle_{0} = 0$. When $\lambda_{\alpha}$ is large, $\langle \hat{n}_{\alpha} \rangle_{N}$ is close to $1$ and the right side of Eq.~\eqref{Eqn:RecursionForN} may develop numerical round-off errors. To avoid this, we reverse the recursion and compute the ``hole'' distribution function for large values of $\lambda_{\alpha}$
\begin{equation}
    1 - \langle \hat{n}_{\alpha} \rangle_{N} = \frac{e^{\beta \mu}P_{\mu}(N + 1)}{\lambda_{\alpha} P_{\mu}(N)} \langle \hat{n}_{\alpha} \rangle_{N + 1}
\end{equation}
with the initial condition $\langle \hat{n}_{\alpha} \rangle_{N_s} = 1$ at unit filling.

The expectation value of the two-body density operator can be computed in a similar fashion:
\begin{eqnarray}
    \langle \hat{c}^{\dag}_i \hat{c}_j \hat{c}^{\dag}_k \hat{c}_l \rangle_{N} &=& \sum_{\alpha, \beta} [ \textbf{P}_{i \alpha} \textbf{P}_{k \beta} \langle \hat{n}_{\alpha} \hat{n}_{\beta} \rangle_{N} \textbf{P}^{-1}_{\beta l} \textbf{P}^{-1}_{\alpha j} \nonumber \\
    &+& \textbf{P}_{i \alpha} \textbf{P}_{k \beta} \langle \hat{n}_{\alpha} - \hat{n}_{\alpha} \hat{n}_{\beta} \rangle_{N} \textbf{P}^{-1}_{\beta j} \textbf{P}^{-1}_{\alpha l}]
\end{eqnarray}
and the two-level correlations can be expressed as 
\begin{equation}
    \langle\hat{n}_{\alpha} \hat{n}_{\beta} \rangle_{N} = 
    \begin{cases} 
    \langle \hat{n}_{\alpha} \rangle_{N} & \alpha = \beta, \\
    \\
    \frac{\lambda_\beta \langle \hat{n}_{\alpha} \rangle_{N} - \lambda_\alpha \langle \hat{n}_{\beta} \rangle_{N}}{\lambda_\beta - \lambda_\alpha} & \alpha \neq \beta,
    \end{cases}
\end{equation}
which can be viewed as the lowest order canonical ensemble generalization
\cite{schonhammer2017deviations,Giraud:2018wt} of Wick's theorem
\cite{Wick:1950wt}, recently extended to the case of degenerate spectra \cite{barghathi2020theory}. These expressions are used throughout the rest of the paper to compute energies, particle densities, and correlation functions. 

\subsection{Model System: Fermion Hubbard Model}

While our formalism generalizes to any two-body Hamiltonian, for the sake of subsequent discussion, we will focus on the fermion Hubbard model due to the strong correlation it exhibits and its relevance for the description of many useful chemical and material systems. This model's Hamiltonian may be expressed as
\begin{equation}
\hat{H} = -t \sum_{ij,\sigma} \left(\hat{c}_{i,\sigma}^{\dagger} \hat{c}_{j,\sigma}^{\phantom \dagger} + \text{H.c.}\right) + U \sum_{i} \hat{n}_{i,\uparrow} \hat{n}_{i,\downarrow}
\end{equation}
where $\hat{c}_{i,\sigma}^{\dagger}(\hat{c}_{j,\sigma}^{\phantom\dagger})$ are anti-commuting fermionic creation(anihilation) operators such that $\hat{c}_i^{\phantom \dagger}\hat{c}_j^\dagger + \hat{c}_i^{\dagger}\hat{c}_j^{\phantom\dagger} = \delta_{ij}$ and  $\hat{n}_{i,\sigma} = \hat{c}_{i,\sigma}^{\dagger} \hat{c}_{i,\sigma}^{\phantom \dagger}$ is the local spin-resolved density for hopping parameter $t$ and interaction strength $U$. In our subsequent illustrations, we pose different challenges to our formalism by varying the strength of the electron correlation, $U/t$, the filling (average number of electrons per site), $\langle n \rangle = (\sum_{i}\hat{n}_{i\uparrow} + \hat{n}_{i\downarrow})/N_{s}$, and number of sites, $N_{s}$, in our model. We measure energies in units of the hopping parameter $t$ in our simulations and set $t=1$ in the remainder of this work.

\section{Results and Discussion \label{results}}

All code, scripts, and data needed to reproduce the results in this paper are available online \cite{papers_code_repo,Shen_src}. 

\subsection{Accuracy, Stability, and Speed of Our Recursive Approach \label{accuracy}}

In order to assess the utility of our approach, we begin by characterizing its accuracy, stability, and relative speed. Because our APF-based algorithm no longer involves taking the difference between products of probabilities or partition functions, we expect it to be substantially more stable at low temperatures, enabling us to model larger systems closer to their ground states. While we have previously demonstrated this increased stability for non-interacting spin models \cite{barghathi2020theory}, here we demonstrate that this algorithm is equally stable for interacting systems.      

To illustrate the stability of our algorithm for a non-interacting system directly relevant to our final interacting simulations, we compute the canonical partition functions and occupations of a version of the Hubbard model whose solutions can be derived analytically. As described in Section \ref{sec:afqmc}, in order to fully account for the two-body interactions in the AFQMC formalism, one employs the Hubbard-Stratonovich Transformation to integrate over many possible random instantiations of non-interacting systems parameterized by different auxiliary fields. While the partition function of a general, interacting system is not analytically solvable, we can compute the partition function for a system parameterized by a single set of fields as a function of $N$ and compare its results to those from the projection algorithm. 

\begin{figure}[t]
    \centering
    \includegraphics[width=\linewidth]{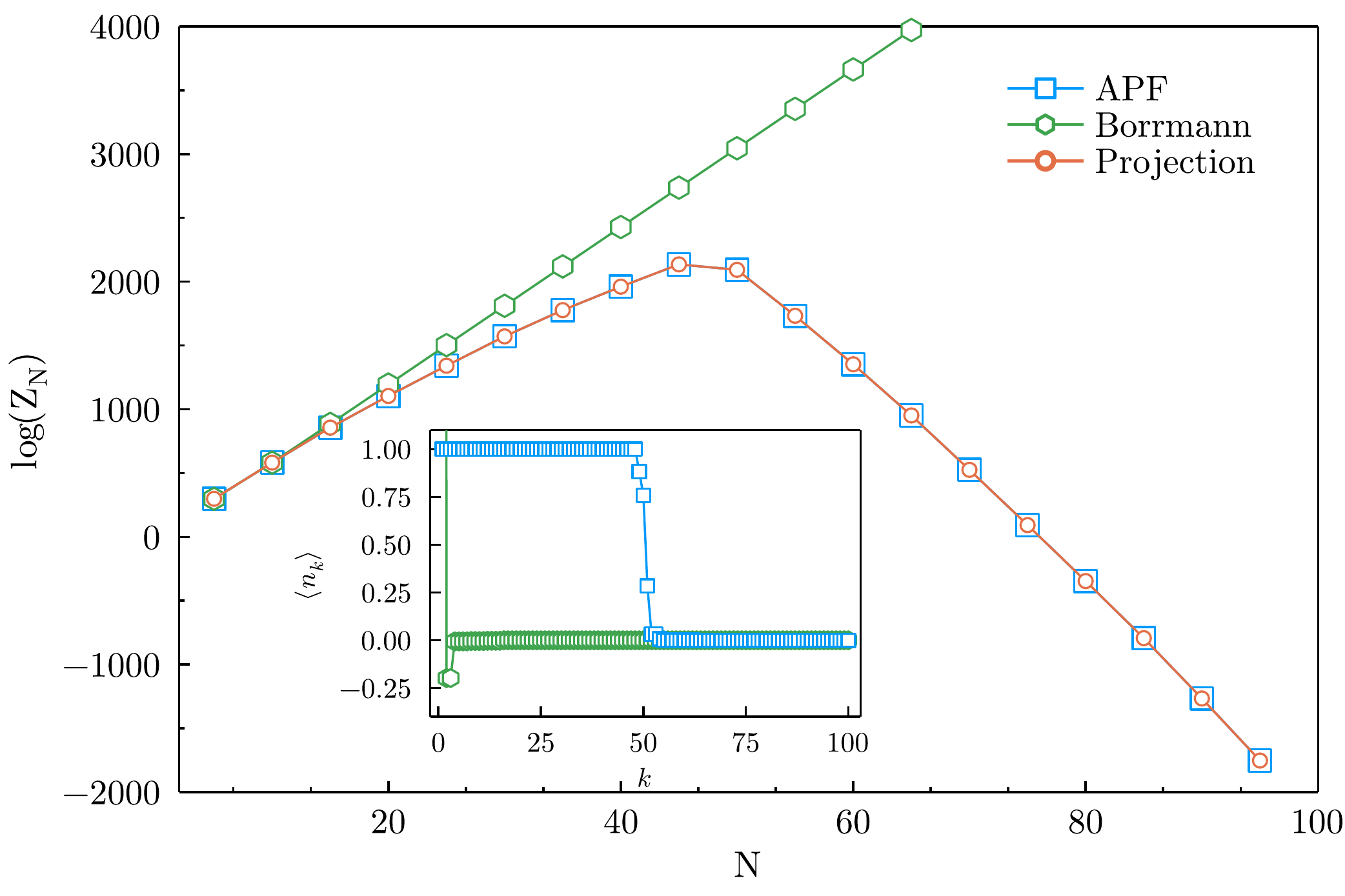}
    \caption{Stability with which different canonical ensemble algorithms can compute the partition function of a 100-site Hubbard chain with a randomly-generated set of Hubbard-Stratonovitch fields for $\beta=10$ for different total particle numbers $N$. The inset depicts the occupation number of all of the energy levels at half-filling ($N=50$). 
    }
    \label{fig:LogZ}
\end{figure}

In Fig.~\ref{fig:LogZ}, we plot the logarithm of the canonical partition function vs.\@ the number of electrons computed using the Borrmann, APF, and projection algorithms for a $100$-site Hubbard chain with a randomly-generated, but known set of Hubbard-Stratonovich fields at $\beta=10$. 
As is evident from the scale on the plot, this is a stringent test of the stability of these algorithms because of the exceedingly large and small values of $Z_{N}$ that can be assumed in this model. Due to binomial combinatorics, the partition function can be expected to peak at half-filling. This is correctly captured by the APF and Projection algorithms, but not by the Bormann recursion. While all algorithms are able to accurately compute the partition function at small fillings, the Borrmann recursion quickly loses its stability (and therefore accuracy) relative to the other algorithms at larger fillings. As previously observed \cite{Shen_JCP_2020}, this is because the Borrmann recursion relies on sums overs terms with alternating signs that can cancel each other out. The instability of the Borrmann algorithm can similarly be observed in recursions for the average occupations for all of the different energy levels, $k$, as presented in the inset of Fig.~ \ref{fig:LogZ}. Here, the Borrmann recursion is not only unable to reproduce the expected non-interacting Fermi-Dirac-like distribution, but even predicts unphysical negative occupations. In contrast, both the APF and Projection algorithms agree regardless of filling, providing clear evidence that the APF formalism is highly stable for non-interacting systems. Given that the interacting partition function may be obtained by integrating over such non-interacting partition functions, the same stability observed in these simulations of non-interacting instantiations of the Hubbard model should naturally extend to simulations of the fully interacting model, as presented below.

Having demonstrated that both the APF and Projection algorithms are highly stable, we next compare their relative computational scaling. Both algorithms make heavy use of full matrix diagonalizations at an $O(N_s^3)$ cost.  However, due to its use of a Fourier sum, the Projection algorithm sums over $N_s$ Fourier components with each component computing a Fourier-frequency-dependent partition function of a single-particle space of size $N_s$. In contrast, the APF algorithm only requires $N$ iterations, where each iteration involves computing the occupation probability of $N_s$ single-particle levels. Ultimately, this results in the recursive algorithm having an $O(N_s^3+N_sN)$ scaling for computing partition functions and occupation numbers (one-body densities), whereas the Projection algorithm has an $O(N_s^3+N_s^2)$ scaling. Thus, there is a clear benefit to using the APF method for filling fractions below unity. The two algorithms also possess different prefactors: the Projection algorithm requires an additional rescaling step to avoid numerical overflow when the values of the Fourier components exceed the available floating point maximum at a given precision, while the APF algorithm is automatically stabilized as the calculations are mapped to probabilities that take values in the range $[0, 1]$. This points to the recursive algorithm being more efficient, particularly at low fillings and for large system sizes.

\begin{figure}[t]
    \centering
    \includegraphics[width=\linewidth]{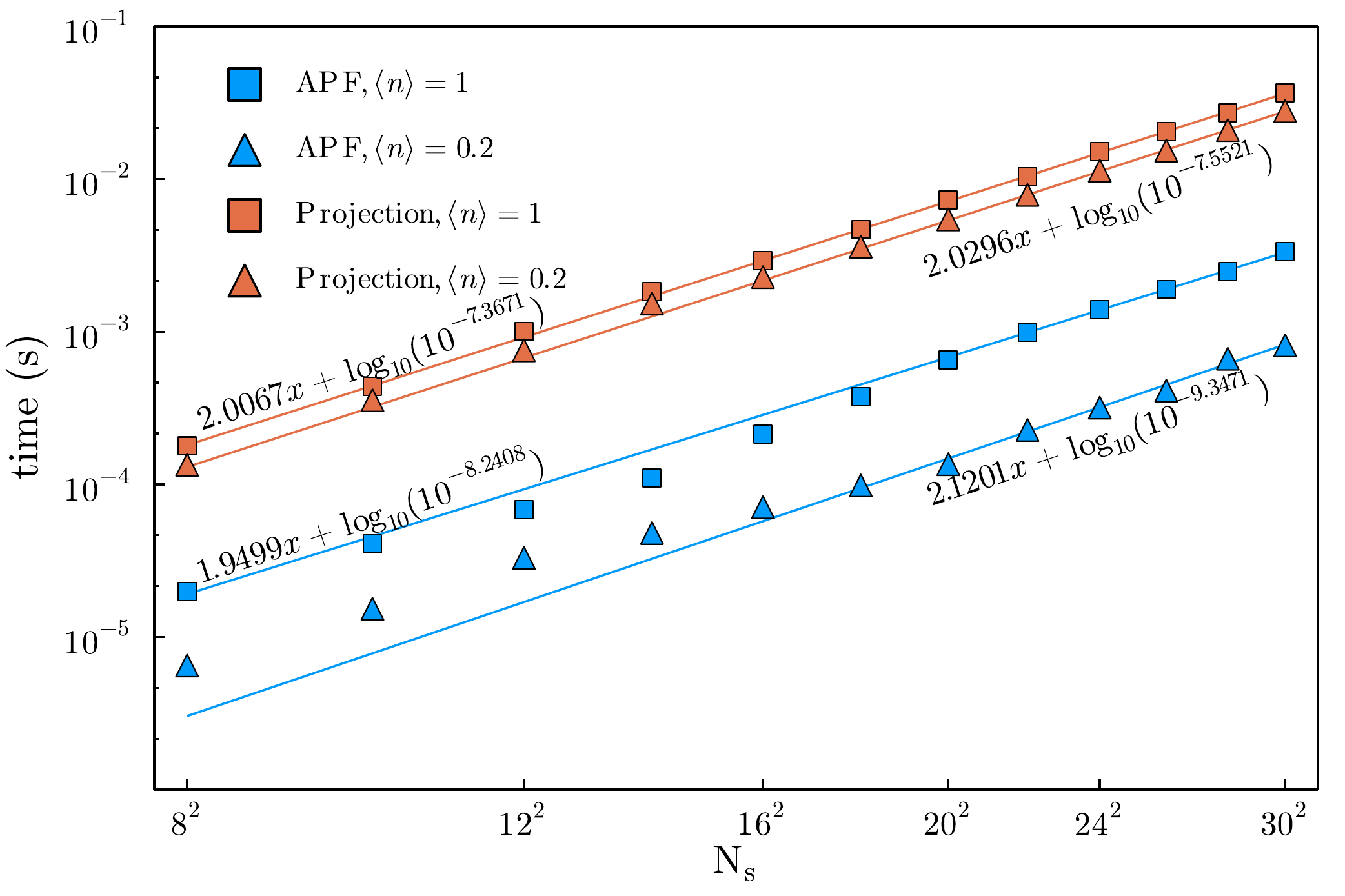}
    \caption{Comparison of the runtime per sample (smaller is better) for computing the partition function of our Auxiliary Partition Function (APF) algorithm against the more conventional Projection algorithm for Hubbard models with varying numbers of lattice sites, $N_{s}$ with $U=2$ at two different fillings $\expval{n} = 1$ and $\expval{n} = 0.2$, and $\beta=12$. Runtimes at $\expval{n} = 1$ are denoted by the squares, while those for the $\expval{n} = 0.2$ filling are denoted by triangles.}
    \label{fig:Runtime}
\end{figure}

The increased efficiency of the recursive algorithm is reflected in Fig.~ \ref{fig:Runtime}, which shows the wall time required to compute the canonical partition function, $Z_N(\sigma)$, of a random sample drawn from a HS-transformed Hubbard model with $U=2$ and $\beta=12$ for varying numbers of sites. 
The $O(N_s^3)$ diagonalization step is identical for both methods and is not included here to foster a direct comparison.  Regardless of the size of the system and the filling, we find that our APF algorithm is always faster than the Projection algorithm, in line with our scaling derivations. As expected, we also observe that the time to run the Projection algorithm remained roughly the same for different filling fractions, while the wall time of the APF algorithm significantly decreased at lower fillings. Performing linear regression on the log-log data in Fig.~\ref{fig:Runtime} quantifies the overall $N_s^2$ scaling, (regression slopes), and reduced APF prefactor, reflected in the regression intercepts, due to the improved algorithmic efficiency of the APF method. 

Although both algorithms could potentially be furthered optimized, we believe that the scalings described here are those representative of typical implementations of these algorithms. As further discussed in the Supplementary Materials \cite{supplemental}, computational complexities can also be worked out for the evaluation of the level occupations and their correlation functions. We find that the cost to compute occupations and related observables follows roughly the same scaling as for calculations of the partition function. 

\subsection{Ground State Convergence Properties}

\subsubsection{Convergence of the Energy to the Ground State}

Having demonstrated the markedly improved stability of our new method, we can now not only assess how it performs on fully interacting systems, but contrast the different physics that emerges in the canonical vs.\@ the grand canonical ensemble down to relatively low temperatures. To appreciate these disparities, we begin by comparing how the energies of the Hubbard model converge to their ground state energies at fixed $N$ and $\mu$. The two ensembles are most effectively compared by choosing a grand canonical $\mu$ such that the average number of particles is given by $\langle{\hat{N}\rangle}_\mu = N$.

In the past, the convergence of the energy to its ground state value with decreasing temperature has commonly been used to assess the relative contribution of thermal quasiparticle excitations above the ground state to the overall state of the quantum system. 
These excitations are also a key contributor to the free energy, the key property describing finite temperature thermodynamics.

\begin{figure}[t]
    \centering
    \includegraphics[width=\linewidth]{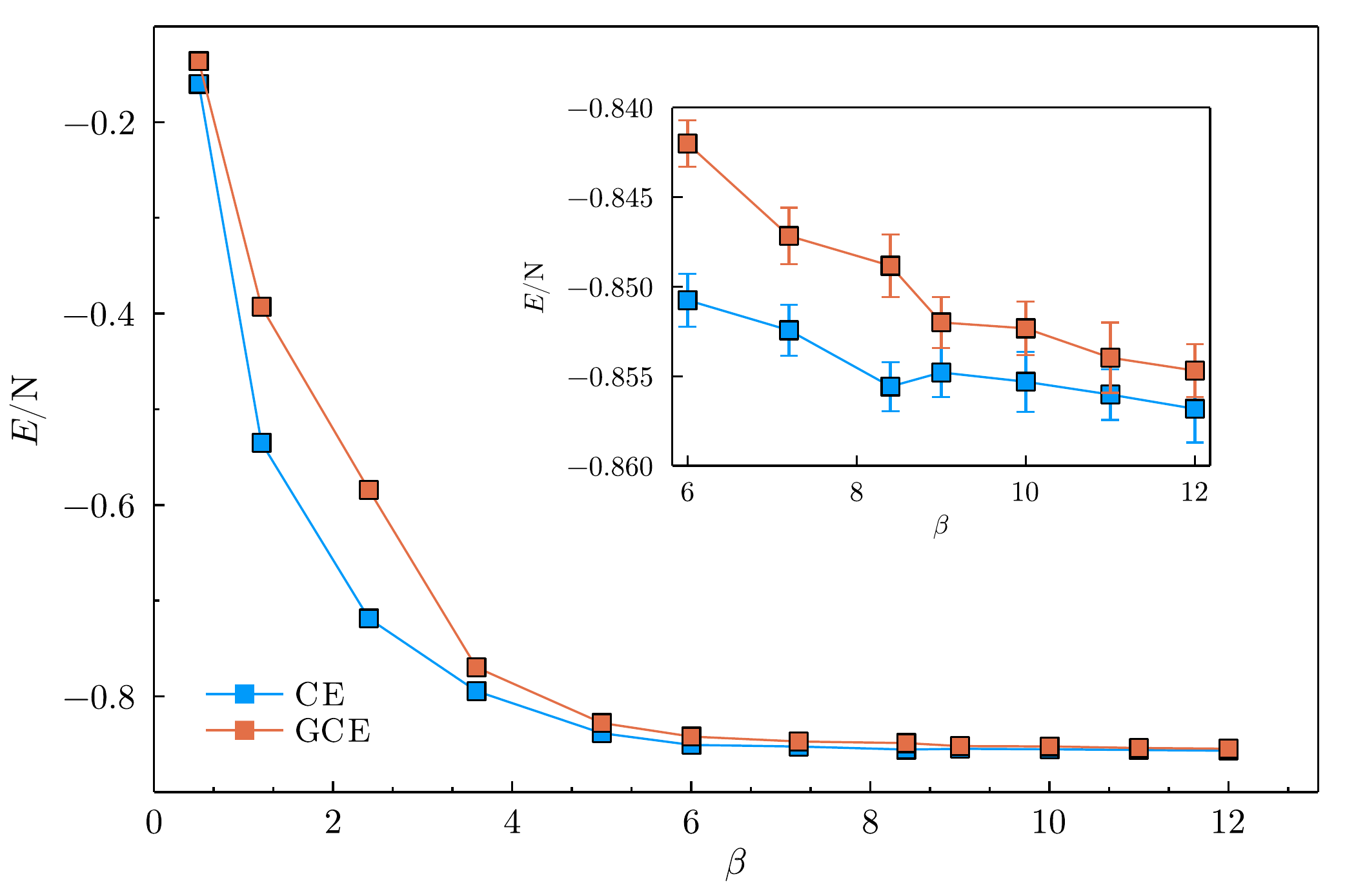}
    \caption{Convergence of the energy per electron in the canonical (CE) and grand canonical (GCE) ensembles as a function of the inverse temperature for a 6$\times$6 Hubbard model at half-filling with $U=4$. Both the main panel and inset demonstrate the enhanced convergence in the canonical ensemble.} \label{fig:EnergyConvergence_6x6}
\end{figure}

In Fig.~\ref{fig:EnergyConvergence_6x6}, we plot the energy per electron vs.\@ the inverse temperature for the $6\times 6$ Hubbard model at half-filling with $U=4$ (and $t=1$). 
Both the canonical and grand canonical ensembles yield predictably large energies at high temperatures due to the higher thermal energy enabling the electrons to access higher energy states, and then converge to roughly the same energy at lower temperatures (large $\beta$). For all values of $\beta$, the canonical ensemble energy is lower, with the largest difference occurring at intermediate temperatures. This is because a larger number of higher energy states are accessible to the electrons in the grand canonical ensemble than in the canonical ensemble due to number fluctuations. This difference between ensembles grows with increased filling and decreased lattice size, which is in line with the intuition that differences between the ensembles should decrease as the thermodynamic limit is approached. Interestingly, despite the differences in their state spaces, both ensembles appear to converge to the same ground state energy on the scale of Fig.~\ref{fig:EnergyConvergence_6x6} at low temperatures. As we shall show next, the energy turns out to be too blunt of a metric to detect subtle and potentially important differences between ensembles.

\subsubsection{Convergence of the Purity and Fidelity to the Ground State}

Given the similar convergence of the energy to the ground state in both ensembles discussed in the previous section, one may ask if there are more fundamental metrics sensitive to differences between the two ensembles. Indeed, the two ensembles are comprised of different states that are accessible at different temperatures which should lead to differences in their convergence to the ground state. 

For any finite-sized system, there exists a crossover temperature below which the system is effectively in its ground state.  This can serve as a more direct indicator for comparing the convergence rate between ensembles than by directly comparing the $\beta$-dependence of the total energy. To quantitatively determine the crossover temperature, we are inspired by information theory and turn to the measurement of the purity
\begin{equation}
    \mathcal{P} = \Tr \hat{\rho}^2, \label{eqn:Purity}
\end{equation}
where $\hat{\rho}$ is the thermal density matrix. The purity quantifies the how mixed a given finite temperature state is: in general, any finite temperature state is mixed and cannot be represented as a single vector in Hilbert space, resulting in a purity of less than 1, $\mathcal{P} < 1$. Thus, quantitative deviations of the purity from the identity can provide insights into the convergence to the ground state in a more general fashion than investigating any individual physical quantity whose $T=0$ value may not be known in general.

The purity can be computed in QMC through a replica trick \cite{Calabrese:2004ll,Daley:2012rt} by rewriting it in terms of a ratio of partition functions
\begin{align}
    \Tr \hat{\rho}^2 = \frac{Z(2\beta)}{Z^2(\beta)} &= \frac{\int_{\sigma_1, \sigma_2} \text{Tr} \left(\hat{U}_{\sigma_1} \hat{U}_{\sigma_2} \right)}{\int_{\sigma_1, \sigma_2} \Tr \left(\hat{U}_{\sigma_1} \right) \text{Tr} \left(\hat{U}_{\sigma_2} \right)} \nonumber \\
    &= \frac{\int_{\sigma_1, \sigma_2} Z(\sigma_1 \cup \sigma_2)}{\int_{\sigma_1, \sigma_2} Z(\sigma_1) Z(\sigma_2)}\, .
    \label{PurityEquation}
\end{align}
Here, $Z(\sigma_1)$ and $Z(\sigma_2)$ are the usual partition functions (in either the canonical or grand canonical ensembles) as a function of their Hubbard-Stratonovich fields $\sigma_1$ and $\sigma_2$. $Z(\sigma_1 \cup \sigma_2)$ can be viewed as the partition function for a connected ensemble propagating from $0$ to $2\beta$. The ensemble switching technique \cite{broecker2014renyi, assaad2015stable} can then be adopted to efficiently sample this ratio of partition functions.

Although there is no known closed form for the purity, one can still expand $\rho^2$ in terms of a system's energy levels, and in the ground state (large-$\beta$) limit, only the leading term remains (see the Supplementary Materials \cite{supplemental} for a full derivation)
\begin{equation}
    \mathcal{P} \propto 1 - 2e^{-\beta \Delta E}\, . \label{Eqn:PurityLowTScaling}
\end{equation}
In the canonical ensemble, $\Delta E$ exactly corresponds to the energy gap between the ground state and first excited state, while in the grand canonical ensemble, $\Delta E$ represents an effective energy gap that contains contributions from states with $N-1$ and $N+1$ particles as sub-leading terms. As a result, a temperature below which the system falls into its ground state region can be revealed by the onset of linear behavior when plotting $\log(1 - \mathcal{P})$ vs.\@ $\beta$.
\begin{figure}[t]
    \centering
    \includegraphics[width=\linewidth]{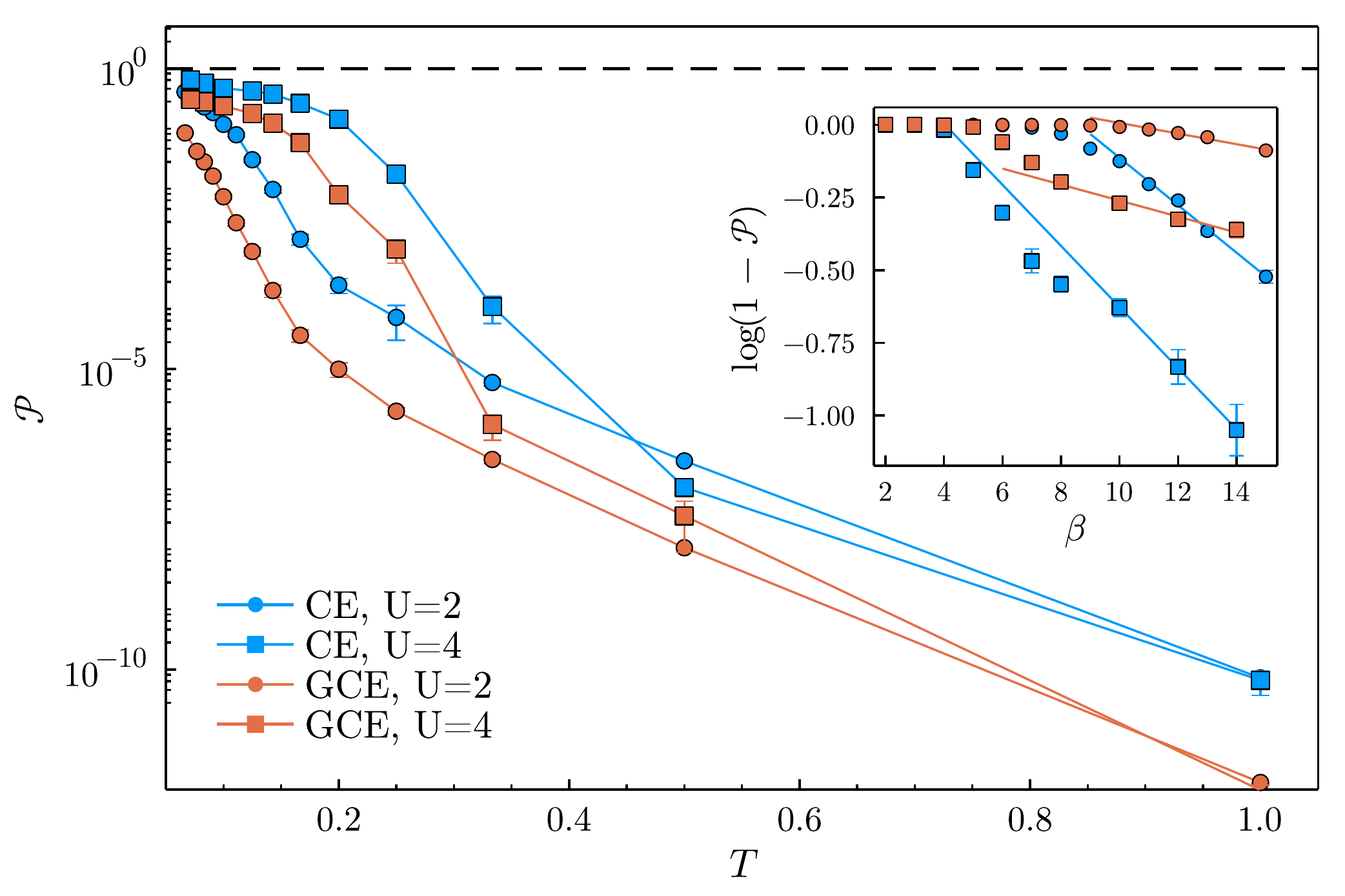}
    \caption{The purity, $\mathcal{P}$, of a 6$\times$6 Hubbard model at half-filling with $U=2$ and $U=4$ as a function of temperature for the canonical (CE) and grand canonical (GCE) ensemble. Different symbols correspond to different interaction strengths and the lines are a guide to the eye. The inset shows $\log(1 - \mathcal{P})$ as a function of the inverse temperature $\beta$, with the crossover to linear behavior (fitted lines) indicating converge to the ground state (see Eq.~\eqref{Eqn:PurityLowTScaling}).}
    \label{fig:Purity}
\end{figure}
In Fig.~\ref{fig:Purity}, we show the purity vs.\@ temperature for Hubbard models simulated in the canonical and grand canonical ensembles for two interaction strengths $U$, and fit $\log(1 - \mathcal{P})$ linearly against the inverse temperature in the low-temperature region in the inset.  As before, simulations performed in the canonical ensemble converge more rapidly to the ground state than those performed in the grand canonical ensemble, as indicated by the uniformly larger canonical purities for any given $U$.  The canonical and grand canonical ensemble purities also show the greatest agreement for $U=4$ for $\beta>6$, when $\log(1 - \mathcal{P})$ begins to significantly deviate below zero, which echoes the convergence for $\beta>6$ seen earlier in the energy. 

However, in contrast to the relative convergence of the energy, the purity reveals additional trends rooted in the underlying physics of the finite temperature crossovers.  In particular, $\log(1 - \mathcal{P})$ can be seen to crossover from exhibiting roughly constant behavior at high temperatures to linear decay with decreasing temperature (increasing $\beta$), as predicted by Eq.~\eqref{Eqn:PurityLowTScaling}. This is a clear indicator that the system has crossed into a regime that is resolving the ground states. By fitting the linear decay of $\log(1 - \mathcal{P})$, we can also extract the energy gap $\Delta E$ from the slope of the regression lines. While it can be statistically challenging to fit curves to such small purity values in the presence of Monte Carlo uncertainties, our fitting procedure yields gaps of $-0.0816$ and $-0.1051$ for the canonical ensemble at $U=2$ and $U=4$, respectively, and $-0.0182$ and $-0.0275$ for the grand canonical ensemble at $U=2$ and $U=4$, respectively. Interestingly, the more negative canonical ensemble slopes suggest that the canonical ensemble gaps are larger than the effective grand canonical gaps, which is supported by the fact that more midgap states may be present in a grand canonical treatment. Moreover, based upon the effective gaps extracted, the systems with $U=4$ possess larger gaps than those with $U=2$, which is expected given the direct correlation between larger $U$ and larger $\Delta E$. A more detailed discussion of the information that can be obtained from the purity is presented in the Supplementary Materials \cite{supplemental}.

Although the purity has provided a more detailed glimpse into the physics of the crossovers that occur in the different ensembles, it is not a measure that directly compares the two ensembles. One metric that can draw such a direct comparison is the fidelity, $\mathcal{F}(\rho, \rho')$, that measures the similarity between two density matrices $\rho$ and $\rho'$. 

In this case, the thermal density matrices in the two ensembles, $\rho_N$ and $\rho_\mu$, describe states that are not pure, and the mixed-state fidelity can be defined as the Hilbert-Schmidt inner product of $\rho_N$ and $\rho_\mu$ normalized by the two purities \cite{WANG2008,Gilles2022_PRB} 
\begin{equation}
    \mathcal{F}(\rho_{N}, \rho_{\mu}) = \frac{\mathrm{Tr}(\rho_{N} \rho_{\mu})}{\sqrt{\mathrm{Tr}(\rho_{N}^2) \mathrm{Tr}(\rho_{\mu}^2)}} \label{Eqn:Fidelity}\, .
\end{equation}
It is straightforward to show that under this definition, the fidelity is normalized and reaches its maximum value of $1$ if and only if $\rho_{N}=\rho_{\mu}$, and that it is also symmetric under the exchange of $\rho_N$ and $\rho_\mu$, $\mathcal{F}(\rho_{N}, \rho_{\mu}) = \mathcal{F}(\rho_{\mu}, \rho_{N})$. Note that the trace in the numerator is taken over the whole Fock space, so the matrix form of $\rho_{N}$ is expanded from the $N$-particle Hilbert space to the Fock space with varying particle numbers, but is only non-zero in the $N$-particle block. Moreover, when the number operator, $\hat{N}$, commutes with the Hamiltonian, as is the case for the Hubbard model, $\rho_{\mu}$ is block-diagonal in the particle number regardless of the interaction strength. This fact allows us to simplify the numerator of Eq.~\eqref{Eqn:Fidelity} to $\mathrm{Tr}(\rho_{N} \rho_{\mu}) = P_{\mu}(N, \beta) \mathrm{Tr}(\rho_{N}^2)$. After some algebra, we arrive at
\begin{eqnarray}
    \mathcal{F}(\rho_{N}, \rho_{\mu}) = \sqrt{P_{\mu}(N, 2\beta)} = \sqrt{\frac{e^{2\beta\mu N} Z_N(2\beta)}{\mathcal{Z}_{\mu}(2\beta)}},
    \label{eq:FdefP}
\end{eqnarray}
which allows $\mathcal{F}(\rho_{N}, \rho_{\mu})$ to be directly measured within our AFQMC simulations, as the ratio between partition functions, $Z_N(2\beta)$ and $\mathcal{Z}_{\mu}(2\beta)$, can be measured through the same ensemble switching technique as was employed in the purity calculations.

In the ground state limit ($\beta \to \infty$), the fidelity has a similar limiting behavior as the purity, which can be derived by expanding the particle-number distribution $P_{\mu}(N, 2\beta)$. A full derivation can be found in the Supplementary Materials \cite{supplemental} and yields
\begin{equation}
    \mathcal{F} \propto 1 - \frac{g}{2} e^{-2\beta \Delta \tilde{E}},
    \label{FidelityLinear}
\end{equation}
where $\Delta \tilde{E}$, is again an effective energy gap that includes the effects of the gap for the system of $N$ particles as its leading term and the gaps for the systems of $N+1$ and $N-1$ particles as its sub-leading terms. $g$ represents an effective degeneracy that accounts for the potentially unresolved spacing of the energy spectrum. This equation implies that plotting $\log(1 - \mathcal{F})$ against $\beta$ is expected to possess linear scaling in the large-$\beta$ limit.

\begin{figure}[t]
    \centering
    \includegraphics[width=\linewidth]{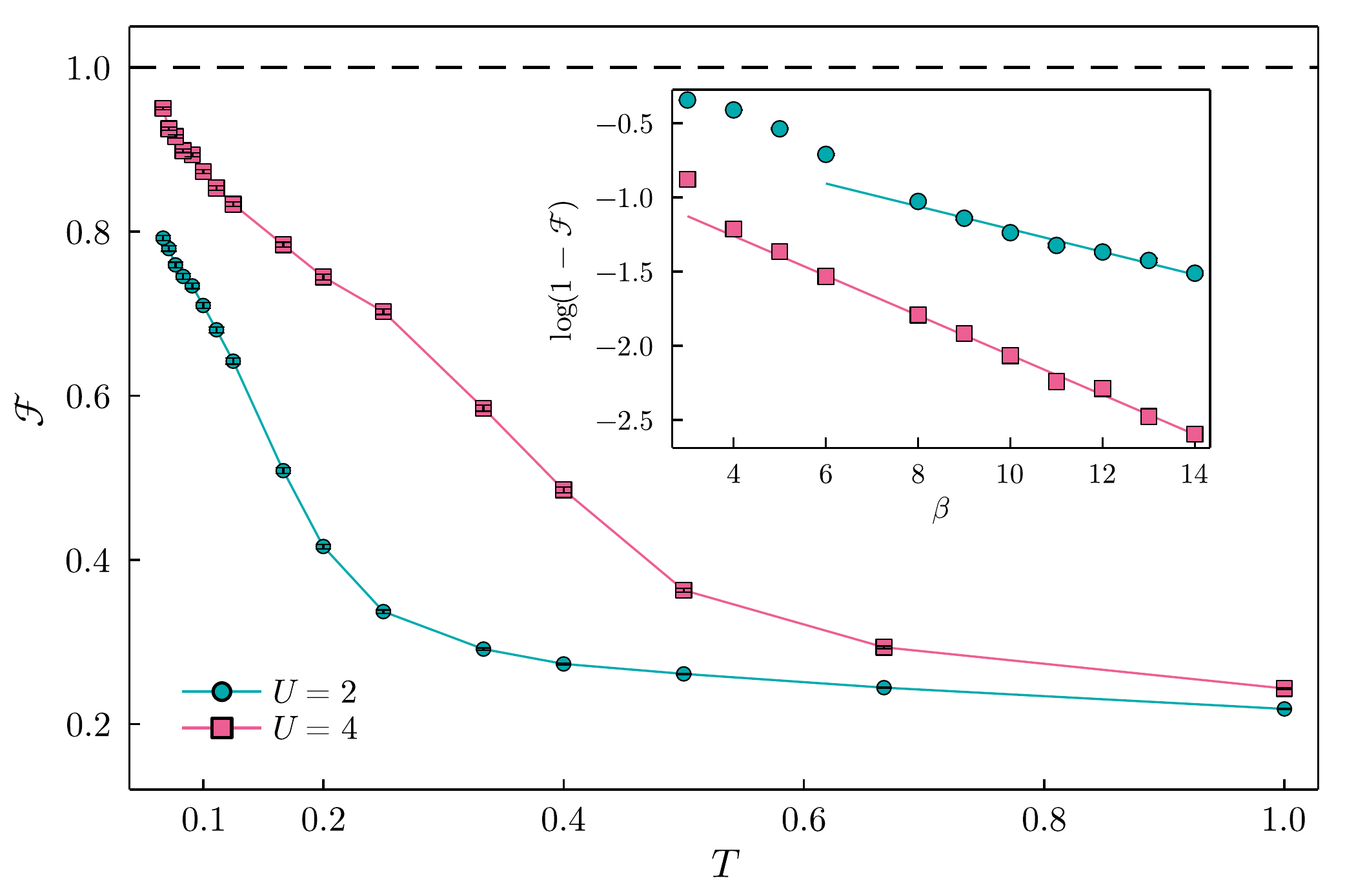}
    \caption{Fidelity as a function of temperature for a 6$\times$6 Hubbard model at half-filling with $U=2$ and $U=4$. The inset fits $\log(1 - \mathcal{F})$ to the inverse temperature from $\beta=8$ to $\beta=14$ and extends to $\beta$ values where $\log(1 - \mathcal{F})$ deviates from a linear fit. The error bars in the inset figure are smaller than the symbols.}
    \label{fig:Fidelity}
\end{figure}

In Fig.~\ref{fig:Fidelity}, we show the fidelity vs. temperature as well as $\log(1 - \mathcal{F})$ vs. $\beta$ in the inset. From this plot, we see that the fidelity is larger for $U=4$ than $U=2$, meaning that, at low temperatures, the ensembles are more similar for larger $U$ values. This is likely because larger $U$ values result in a larger gap, which limits how many additional grand canonical states the system can access beyond those occupied in the canonical ensemble. From the inset, we additionally observe how $\log(1 - \mathcal{F})$ becomes linear in $\beta$ at low temperatures, as predicted by Eq.~\eqref{FidelityLinear}. A larger effective gap is again observed for the more strongly interacting case with $U=4$, which possesses a more negative slope than for $U=2$. The considerable deviation of the fidelity from unity, even at a very low temperature ($T < 0.1$), can be understood from Eq.~\eqref{eq:FdefP}, which definitively captures how particle number fluctuations can continue to contribute to the grand canonical density matrix, even in the limit of large systems when approaching the ground state. 

Although the gaps obtained may not yet be as accurate as those obtained from excited state calculations, these examples illustrate that the purity and fidelity are much more informative metrics of convergence than the energy alone, and provide additional information that can be exploited to estimate gaps from finite temperature simulations. 

\subsection{Sign Problem in the Canonical Ensemble}

After observing how the canonical ensemble converges more rapidly to the ground state, one may ask if this provides a practical way of more readily accessing $\beta \to \infty$ quantities than in the grand canonical ensemble. After all, it is reasonable to assume that if the energy and wave function converge more rapidly in the canonical ensemble, perhaps one can more readily gain access to low-temperature physics before a severe sign problem sets in at certain fillings. 

\begin{figure}[t]
    \centering
    \includegraphics[width=\linewidth]{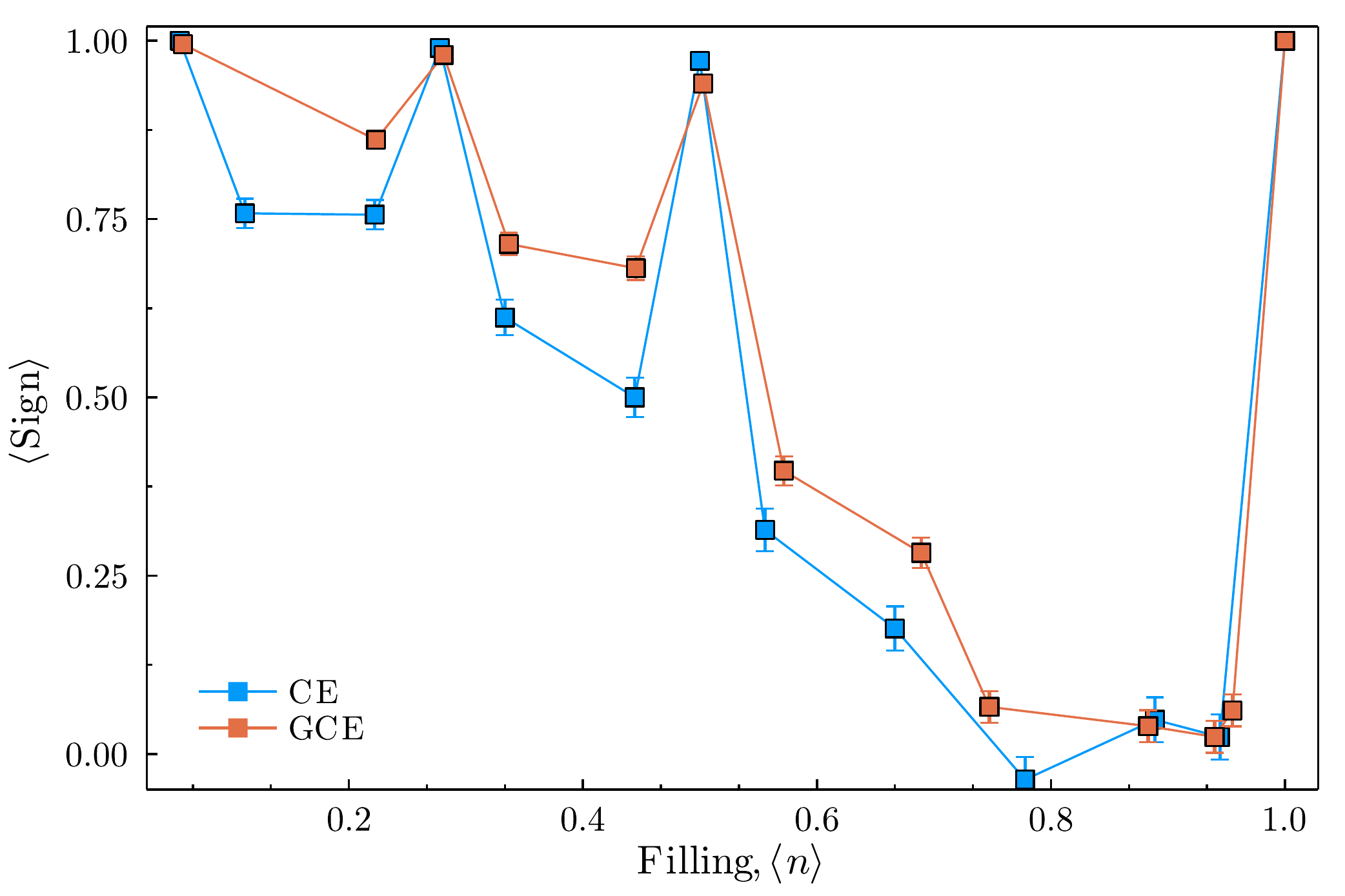}
    \caption{Average sign vs.\@ filling fraction in the canonical (CE) and grand canonical (GCE) ensembles of a 6$\times$6 Hubbard model with $U=4$ and $\beta=10$.}
    \label{fig:SignConvergence_6x612beta10}
\end{figure}

To address the interplay between convergence to the ground state and the emergence of a physical fermion sign problem, we compute the average sign at different fillings for a variety of temperatures and interaction strengths. The behavior of the sign as a function of filling presented in Fig.~\ref{fig:SignConvergence_6x612beta10} is representative of what we more widely observe: in general, the average sign in the canonical ensemble is less than that in the grand canonical ensemble at any given filling. 
Just as in the grand canonical ensemble, the sign at certain fillings is at or near 1, reflecting special symmetries and indicating that the system can be modeled with no approximations at polynomial cost. However, away from these special fillings, the average sign decreases, meaning that either exponentially more samples must be taken to converge average observables or sign mitigation strategies must be employed. Unfortunately, where the sign problem is present, the canonical sign problem appears to be more severe. This is likely a consequence of the fact that the canonical ensemble more quickly converges to the ground state: the same states that lower the canonical energy relative to the grand canonical energy are those that give rise to a more significant sign problem. This presents a practical tradeoff. While the canonical ensemble more rapidly converges to the ground state with decreasing temperature, it does so with an increased sign problem, signifying that simulations in the canonical ensemble do not allow for a way to mitigate costs associated with simulating many-body systems of fermions. The ground state of the fermion Hubbard model at most fillings possesses a significant sign problem; the more rapidly this ground state is approached, the more rapidly a sign problem emerges, regardless of ensemble.

\subsection{Impact on Thermometry: Differences Between Canonical and Grand Canonical Ensemble Observables \label{thermometry}}

While the energy and sign are two of the most commonly measured observables in stochastic many-body simulations, densities and correlation functions enable the most direct comparisons with experiments. One may therefore ask what significant differences may exist between canonical and grand canonical densities. This is not an idle question: many recent cold atom experiments estimate the temperature of their trapped gases assuming that the constituent particles interact according to the grand canonical ensemble \cite{Onofrio2016, Mukherjee:2017mz}. Although this is valid at large particle numbers, many such experiments are performed in a mesoscopic regime in which only a finite number of particles are present. Making this assumption when the thermodynamic limit has not been reached, can lead to inaccurate estimates of the system temperature, resulting in incorrect phase diagrams and potentially unfounded efforts to further reduce temperatures.  

To make clear the effect of choosing the incorrect ensemble on the determination of temperature in a finite closed quantum system, it is illustrative to consider the simplest extreme example of one particle ($N=1$) distributed amongst $N_s = 2$ energy levels $\qty{0,\Delta}$. In the canonical ensemble,
\begin{equation}
    \avg{n_1}_1=\frac{1}{1+e^{-\beta\Delta}}; \qquad \expval{n_2}_1 = e^{-\beta
    \Delta} \expval{n_1}_1\, ,
\end{equation}
while in the grand canonical ensemble, these average occupations depend on the chemical
potential $\mu$,
\begin{equation}
    \avg{n_1}_{\mu}=\frac{1}{1+e^{-\beta\mu}}; \qquad 
    \avg{n_2}_{\mu}=\frac{e^{-\beta(\Delta-\mu)}}{1+ e^{-\beta(\Delta-\mu)}}.
    \label{eq:n1n2GC}
\end{equation}
In Eq.~\eqref{eq:n1n2GC}, $\mu$ is chosen such that $\expval{n_1}_{\mu } + \expval{n_2}_{\mu} = 1$, which yields $\mu = \Delta/2$. Using this value in Eq.~\eqref{eq:n1n2GC} and equating occupation numbers between ensembles (the quantity most directly accessible in thermometry experiments through the velocity distribution) would require using an inverse temperature in the grand canonical ensemble that is twice that of the physical canonical temperature, \emph{i.e.}\@ $\beta_{\mu} = 2\beta \equiv 2\beta_N$. This would result in a 100\% error in the extracted temperature (when the incorrect ensemble is chosen), with grand canonical simulations always predicting a \emph{lower} temperature than the physical one.

\begin{figure}[t]
    \centering
    \includegraphics[width=\linewidth]{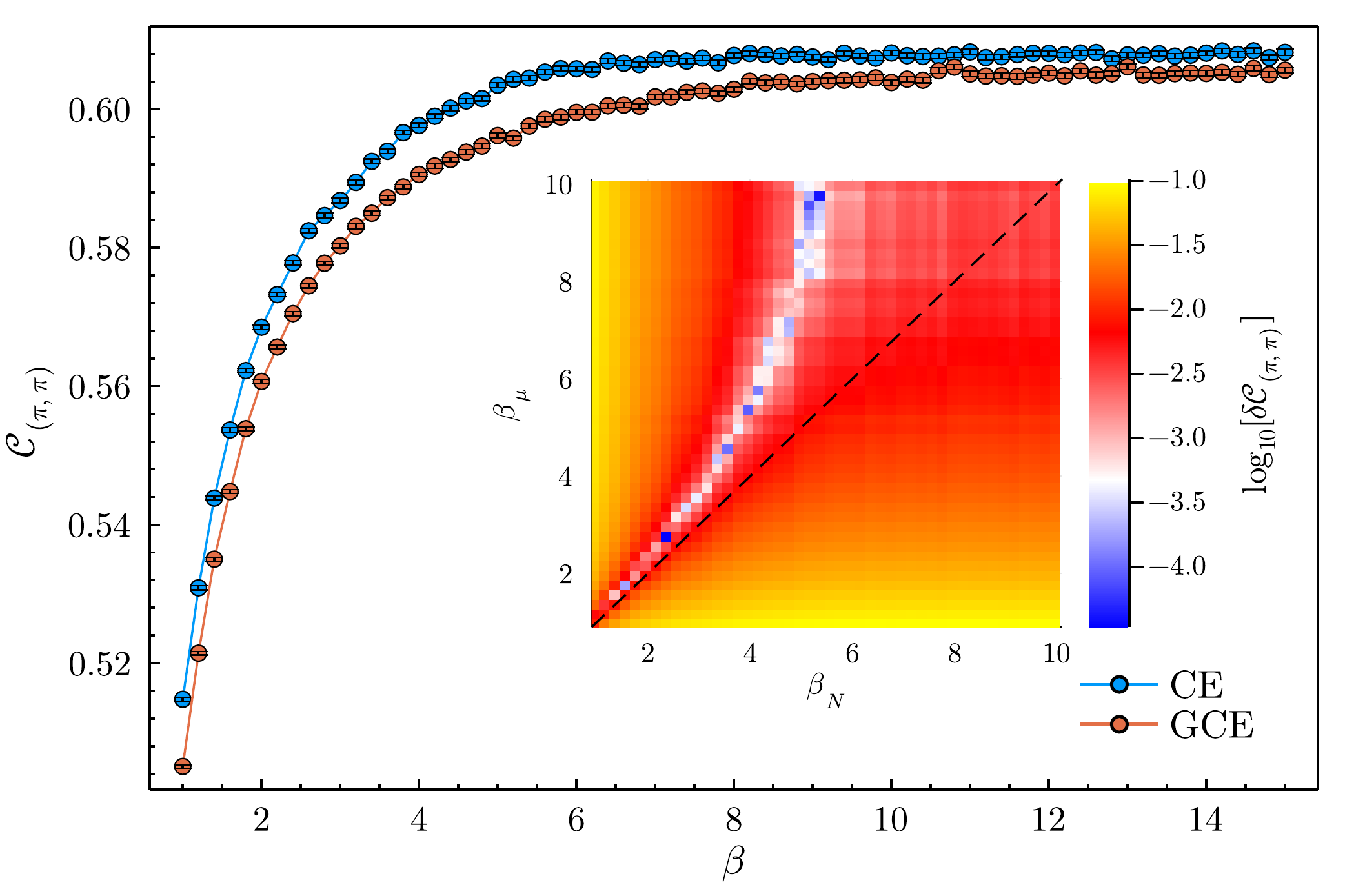}
    \caption{Charge structure factor at $\textbf{k}=(\pi, \pi)$,
    $\mathcal{C}_{(\pi, \pi)}$, as a function of inverse temperature $\beta$ in
the canonical (CE) and grand canonical (GCE) ensemble for a $6\times 6$ Hubbard model at $U=2, \obs{n} = \frac{2 \times 23}{36} \approx 1.28$. The inset shows the heatmap of the difference in the
charge structural factors, $\delta\mathcal{C}_{(\pi, \pi)}(\beta_{N},
\beta_{\mu})$, as a function of the inverse temperature in the canonical and
grand canonical ensembles.} 
    \label{fig:Cq}
\end{figure}

While this is obviously an extreme (toy) example, thermometry errors can persist to
larger systems that include interactions that affect other measurable quantities.  
For example, second-order fluctuations in the particle number provide an even
clearer way to explore the impact of choosing the incorrect
ensemble. This can be quantified by the site occupancy correlation function
\begin{equation}
    \obs{\hat{n}_{\textbf{i}} \hat{n}_{\textbf{j}}} = \obs{(\hat{n}_{\textbf{i}, \uparrow} + \hat{n}_{\textbf{i}, \downarrow})(\hat{n}_{\textbf{j}, \uparrow} + \hat{n}_{\textbf{j}, \downarrow})},
    \label{eq:ninj}
\end{equation}
where $\textbf{i},\textbf{j}$ are $D$-dimensional site indices. This correlation function is more naturally studied in momentum space via the static charge structure factor at wavevector $\textbf{k}$,  given by the Fourier transform of
Eq.~\eqref{eq:ninj}:
\begin{equation}
    \mathcal{C}_{\textbf{k}} = \frac{1}{N_s} \sum_{\textbf{i}, \textbf{j}} e^{\imath \textbf{k}
    \cdot (\textbf{i} - \textbf{j})} \obs{\hat{n}_{\textbf{i}}
\hat{n}_{\textbf{j}}},
\end{equation}
which is, in principle, measurable in cold atom experiments through, \emph{e.g.}\@ Bragg
spectroscopy \cite{Altman:2004aa,Kuhnle:2010os,Boll:2016oa,Yang:2018po}. To
quantify the difference between ensemble predictions for such a quantity, we
compute $\mathcal{C}_{\vb{k}}$  at $\vb{k} = (\pi,\pi)$ to expose antiferromagnetic correlations for a 2D Hubbard model at $U=2$ and filling
$\expval{n} = \frac{2 \times 23}{36}$ as a function of inverse temperature $\beta$ as shown in Fig.~\ref{fig:Cq}. 
The odd number of electrons was chosen due to the existence of the large gap and the increased spacing of low-lying energy levels.The grand canonical ensemble simulations off of half-filling are implemented by tuning the chemical potential dynamically \cite{Miles:2022hr} during the Monte Carlo step to ensure $\abs{\expval{n}_\mu - 1.28} < 1e^{-3}$. The deviation at each temperature is clear, with the canonical structure factor having a consistently larger value due to suppressed fluctuations. The inset shows a heat-map of the difference $\delta C_{\vb{k}}$ for each temperature $\beta \equiv \beta_N$, with the $y$-axis representing the effective inverse temperature $\beta_\mu$ needed to minimize the difference between ensembles (the horizontal shift in the main panel needed to align the two curves).  Deviations of the minimum from the line
$\beta_\mu = \beta_N$ captures potential errors in thermometry when using the
incorrect ensemble. The maximum deviation in the heatmap over the temperature range we study appears to be $(\beta_N, \beta_\mu) \approx (4.4, 9.4)$, which leads to a 53.2\% thermometry error. Such large errors underscore the dramatic errors that can arise from an incorrect choice of ensemble and that can be addressed using the new techniques presented in this work. 

\section{Conclusions \label{conclusions}}
In summary, we have presented and illustrated applications of a new, significantly more stable recursive algorithm for determining the physics of interacting systems in the canonical ensemble. This algorithm integrates the Auxiliary Partition Function formalism, a highly stable means of computing the properties of non-interacting systems in the canonical ensemble, into the Auxiliary Field Quantum Monte Carlo framework by exploiting the Hubbard-Stratonovich Transformation. We demonstrate the stability of this algorithm and then showcase its potential applications by studying differences in the way that the canonical and grand canonical ensembles converge to the ground state. This convergence is quantified using information theory metrics by comparing the purity of finite temperature states generated within the grand canonical and canonical frameworks.  We find that the canonical AFQMC results in a suppression of the mixed state and improved fidelity with the ground state as $T \to 0$ in a practical simulation.  

As a potentially experimentally relevant application, we show that a grand canonical treatment of the thermometry of cold atom and other systems with fixed particle numbers can lead to underestimates of the temperature of those systems, clouding investigations of their thermodynamics.  This is becoming more pressing as studies of trapped ultra-cold fermions push into a regime of smaller $T/T_F$ where they are more poorly described by the grand canonical ensemble \cite{Hueck:2018td}. Moreover, this work has direct implications for the study of nuclear matter, which possesses fixed nucleon numbers and has traditionally been modeled using the Projection algorithm \cite{Ormand_PRC,Jensen_PRL_2020}. 

The fact that we find our algorithm to be more computationally-efficient than oft-used Projection algorithms opens the door to a wealth of potential new applications. To date, most Projection algorithms have been limited to system sizes of tens, to at the very most, low hundreds of particles, with application to smaller nuclei and ``toy'' condensed matter systems. In its present form, without many algorithmic advances or computational fine-tuning, our algorithm can readily model systems with many hundreds of particles. This opens the door to more accurate numerical descriptions of cold atom quantum simulators or mesoscale  devices where discrete particle number fluctuations can influence the transport of heat and matter \cite{Campisi:2011fr}.

Although we illustrated the performance of our algorithm on systems of fermions because of their greater relevance and potential to develop a sign problem, our algorithm, with appropriate modifications, is equally applicable to systems of bosons or particles with other quantum statistics. The fact that the algorithm does not require explicit knowledge of many-body chemical potentials and is stable for large systems implies that it can see wide application in the study of quantum condensates, which can be challenging to simulate in the grand canonical ensemble \cite{rubenstein2012finite,Gilbreth_PRA_2013}. 

We also anticipate that our algorithm will enable more direct comparisons with other finite temperature algorithms, including the Path Integral Monte Carlo \cite{Hirshberg_PNAS,Malone_PRL_2016,Dornheim_PRL}, Density Matrix Quantum Monte Carlo \cite{Petras_JCTC,Malone_JCP_2015}, and emerging Finite Temperature Coupled Cluster theories \cite{White_Chan_2018,Hirata_Jha_2020}, all of which are formulated in the canonical ensemble. Beyond the algorithmic, our work will furthermore enable seamless canonical ensemble simulations from high temperatures to the ground state where most simulations are inherently performed in the canonical ensemble. This will provide critical insights into how finite temperature physics gives rise to ground state physics in correlated systems with decreasing temperature without the noise induced by spurious particle number fluctuations. Given the correlations that fixed-particle number constraints impose on the occupancies of different states, we moreover anticipate that fluctuations and therefore the physics of systems in the canonical ensemble will be fundamentally different. We look forward to the new such canonical ensemble physics this algorithm will reveal. 

\section{Acknowledgements \label{ack}}
The authors thank Richard Stratt, Christopher Gilbreth, Scott Jensen, and Ben Cohen-Stead for fruitful conversations. T.S., J.Y., and B.R. were funded by NSF CTMC CAREER Award 2046744. T.S. is also grateful for financial support from the Brown Open Graduate Education program. A.D.~acknowledges support from 
the NSF under Grant No. DMR-2041995. This research was conducted using computational resources and services at the Center for Computation and Visualization, Brown University.

\nocite{apsrev42Control}
\bibliographystyle{apsrev4-2}
\bibliography{ref}{}

\ifarXiv
    

\section*{Supplementary material}

# Supplementary Information for "A Stable, Recursive Auxiliary Field Quantum Monte Carlo Algorithm in the Canonical Ensemble: Applications to Thermometry and the Hubbard Model"

Tong Shen,[1] Hatem Barghathi,[2] Jiangyong Yu,[3] Adrian Del Maestro,[2,4] and Brenda M. Rubenstein[1,3,*]

[1]*Department of Chemistry, Brown University, Providence, RI 02912*
[2]*Department of Physics and Astronomy, University of Tennessee, Knoxville, TN 37916*
[3]*Department of Physics, Brown University, Providence, RI 02912*
[4]*Min H. Kao Department of Electrical Engineering and Computer Science,
University of Tennessee, Knoxville, TN 37996, USA*

(Dated: December 16, 2022)

## I. ALGORITHMIC DETAILS: MONTE CARLO SAMPLING AND PROPAGATOR STABILIZATION TECHNIQUES

In this section, we provide details regarding how we sample and stabilize our matrices in our new algorithm. Similar to other auxiliary field-based QMC algorithms [S1–S3], one needs to take special care multiplying the propagators, $\mathbf{B}_\mathbf{L}^{(\sigma)}\mathbf{B}_{\mathbf{L-1}}^{(\sigma)}\cdots\mathbf{B}_\mathbf{1}^{(\sigma)}$, used to compute the traces together as directly multiplying the propagators is numerically unstable at low temperatures due to the existence of extremely large and extremely small singular values [S4, S5]. In order to circumvent this, QR factorization with column-pivoting is used to separate singular values of different magnitudes for series of $\mathbf{B}^{(\sigma)}$s of length $l$:

$$\tilde{\mathbf{B}}_\mathbf{k}^{(\sigma)} = \mathbf{B}_{\mathbf{l+k}}^{(\sigma)}\mathbf{B}_{\mathbf{l+k-1}}^{(\sigma)}\cdots\mathbf{B}_\mathbf{k}^{(\sigma)} = \mathbf{U}_\mathbf{k}^{(\sigma)}\mathbf{D}_\mathbf{k}^{(\sigma)}\mathbf{T}_\mathbf{k}^{(\sigma)}, \quad \text{(S1)}$$

where $\mathbf{U}_\mathbf{k}^{(\sigma)}$ is a unitary matrix, $\mathbf{D}_\mathbf{k}^{(\sigma)}$ is a diagonal matrix whose elements range from very large to very small in magnitude, and $\mathbf{T}_\mathbf{k}^{(\sigma)}$ is a regular matrix with a small condition number. Matrix multiplications using this factorization turn out to be stable.

The MC sampling proceeds by updating the $\mathbf{B}_\mathbf{i}^{(\sigma)}$ sequentially. A full update from $i = 1$ to $i = L$ is called a sweep. Before the sweep, we initialize a string of randomly-generated propagator matrices, $\mathbf{B}_\mathbf{L}^{(\sigma)}\mathbf{B}_{\mathbf{L-1}}^{(\sigma)}\cdots\mathbf{B}_\mathbf{1}^{(\sigma)}$, and compute and store all of the partial factorizations as

$$[\underbrace{\mathbf{U}_{\mathbf{n-1}}^{(\sigma)}\mathbf{D}_{\mathbf{n-1}}^{(\sigma)}\mathbf{T}_{\mathbf{n-1}}^{(\sigma)}}_{\tilde{\mathbf{B}}_\mathbf{n}^{(\sigma)}}, \underbrace{\mathbf{U}_{\mathbf{n-2}}^{(\sigma)}\mathbf{D}_{\mathbf{n-2}}^{(\sigma)}\mathbf{T}_{\mathbf{n-2}}^{(\sigma)}}_{\tilde{\mathbf{B}}_\mathbf{n}^{(\sigma)}\tilde{\mathbf{B}}_{\mathbf{n-1}}^{(\sigma)}}, \ldots, \underbrace{\mathbf{U}_\mathbf{1}^{(\sigma)}\mathbf{D}_\mathbf{1}^{(\sigma)}\mathbf{T}_\mathbf{1}^{(\sigma)}}_{\tilde{\mathbf{B}}_\mathbf{n}^{(\sigma)}\cdots\tilde{\mathbf{B}}_\mathbf{2}^{(\sigma)}}]. \quad \text{(S2)}$$

Here, we denote the total number of groups of matrices as $n = L/l$. When updating the $k$th group of matrices, $\tilde{\mathbf{B}}_\mathbf{k}^{(\sigma)}$, in order to reduce the number of QR factorizations, a cyclicly-permuted string of propagator matrices is computed

$$\underbrace{\mathbf{U}_\mathbf{L}\mathbf{D}_\mathbf{L}\mathbf{T}_\mathbf{L}}_{\tilde{\mathbf{B}}_\mathbf{n}^{(\sigma)}\cdots\tilde{\mathbf{B}}_{\mathbf{k+1}}^{(\sigma)}}\tilde{\mathbf{B}}_\mathbf{k}^{(\sigma)}\underbrace{\mathbf{U}_\mathbf{R}\mathbf{D}_\mathbf{R}\mathbf{T}_\mathbf{R}}_{\tilde{\mathbf{B}}_{\mathbf{k-1}}^{(\sigma)}\cdots\tilde{\mathbf{B}}_\mathbf{1}^{(\sigma)}} \to \underbrace{\mathbf{U_C}\mathbf{D_C}\mathbf{T_C}}_{\mathbf{U_R}\mathbf{D_R}\mathbf{T_R}\mathbf{U_L}\mathbf{D_L}\mathbf{T_L}}\tilde{\mathbf{B}}_\mathbf{k}^{(\sigma)}, \quad \text{(S3)}$$

which preserves the value of the trace because the canonical trace is invariant under cyclic permutations, i.e., $\text{tr}_N[\mathbf{B}_\mathbf{L}^{(\sigma)}\cdots\mathbf{B}_\mathbf{1}^{(\sigma)}] = \text{tr}_N[\mathbf{U_C}\mathbf{D_C}\mathbf{T_C}\tilde{\mathbf{B}}_\mathbf{k}^{(\sigma)}]$. As we will show in the following, this form reduces the number of QR factorizations needed for each update to two. Since the left factorization $\mathbf{U_L}\mathbf{D_L}\mathbf{T_L}$ has been precomputed and stored in Eq.(S2), it can be directly read as $\mathbf{U_L}\mathbf{D_L}\mathbf{T_L} = \mathbf{U}_\mathbf{k}^{(\sigma)}\mathbf{D}_\mathbf{k}^{(\sigma)}\mathbf{T}_\mathbf{k}^{(\sigma)}$, and only one QR factorization is required to merge two factorizations, $\mathbf{U_C}\mathbf{D_C}\mathbf{T_C} = \mathbf{U_R}\mathbf{D_R}\mathbf{T_R}\mathbf{U_L}\mathbf{D_L}\mathbf{T_L}$.

Every time slice in $\tilde{\mathbf{B}}_\mathbf{k}^{(\sigma)}$ is then updated sequentially according to the Metropolis criterion. That is, we propose a change to the $i$th time slice $\mathbf{B}_\mathbf{i}^{(\sigma_i)}$ in $\tilde{\mathbf{B}}_\mathbf{k}^{(\sigma)}$ and accept the proposal with probability

$$p_i = \min(1, \frac{Z_N(\{\sigma\backslash\sigma_i\}\cup\sigma'_i)}{Z_N(\sigma)}), \quad \text{(S4)}$$

where

$$Z_N(\sigma) = \text{tr}_N[\mathbf{U_C}\mathbf{D_C}\mathbf{T_C}\tilde{\mathbf{B}}_\mathbf{k}^{(\sigma)}]$$
$$= \text{tr}_N[\mathbf{U_C}\mathbf{D_C}\tilde{\mathbf{T}}], \quad \text{(S5)}$$
$$Z_N(\{\sigma\backslash\sigma_i\}\cup\sigma'_i) = \text{tr}_N[\mathbf{B}_{\mathbf{i+1}}^{(\sigma_{i+1})}\cdots\mathbf{B}_{\mathbf{l+k}}^{(\sigma_{l+k})}\mathbf{U_C}\mathbf{D_C}\mathbf{T_C}$$
$$\cdot\mathbf{B}_\mathbf{k}^{(\sigma_\mathbf{k})}\cdots\mathbf{B}_\mathbf{i}^{(\sigma'_\mathbf{i})}]$$
$$= \text{tr}_N[\tilde{\mathbf{U}}\mathbf{D_C}\tilde{\mathbf{T}}]. \quad \text{(S6)}$$

Once $\tilde{\mathbf{B}}_\mathbf{k}^{(\sigma)}$ is updated to $\tilde{\mathbf{B}}_\mathbf{k}^{(\sigma')}$, we update the $k$th partial factorization of Eq.(S2) as

$$\mathbf{U}_\mathbf{k}^{(\sigma)}\mathbf{D}_\mathbf{k}^{(\sigma)}\mathbf{T}_\mathbf{k}^{(\sigma)} \leftarrow \mathbf{U}_\mathbf{k}^{(\sigma')}\mathbf{D}_\mathbf{k}^{(\sigma')}\mathbf{T}_\mathbf{k}^{(\sigma')} = \tilde{\mathbf{B}}_\mathbf{k}^{(\sigma')}\mathbf{U_R}\mathbf{D_R}\mathbf{T_R}. \quad \text{(S7)}$$

Hence, one effectively subtracts one group of matrices from the left factorization, updates this group, and then multiplies it into the right factorization. For the $k$th update, the left factorization can be directly read from the $k$th element in Eq.(S2) and the right factorization is computed at the end of the $(k-1)$th update according

[*] Author to whom correspondence should be addressed: brenda_rubenstein@brown.edu.

to Eq.(S7). Note that at $k = 1$, we have $\mathbf{U_R D_R T_R = I}$. As a result, a total number of $2n$ QR factorizations are performed during each sweep.

After a full sweep is completed, the partial factorization vector now stores all intermediate right factorizations

$$[\underbrace{\mathbf{U}^{(\sigma')}_{n-1}\mathbf{D}^{(\sigma')}_{n-1}\mathbf{T}^{(\sigma')}_{n-1}}_{\tilde{\mathbf{B}}^{(\sigma')}_{n-1}\cdots\tilde{\mathbf{B}}^{(\sigma')}_{1}}, \underbrace{\mathbf{U}^{(\sigma')}_{n-2}\mathbf{D}^{(\sigma')}_{n-2}\mathbf{T}^{(\sigma')}_{n-2}}_{\tilde{\mathbf{B}}^{(\sigma')}_{n-2}\cdots\tilde{\mathbf{B}}^{(\sigma')}_{1}}, \ldots, \underbrace{\mathbf{U}^{(\sigma')}_{1}\mathbf{D}^{(\sigma')}_{1}\mathbf{T}^{(\sigma')}_{1}}_{\tilde{\mathbf{B}}^{(\sigma')}_{1}}]. \tag{S8}$$

In order to reuse the stored factorizations, we need to sweep all of the fields in reverse order in the next run, i.e., we start the sweep from the $k = n$th group of matrices, subtract the $k$th group from the right factorization for the $k$th update, whose value can be read from the $k$th element in Eq.(S8), update this group, and then multiply it into the left factorization and store the results to the partial factorization vector as in Eq.(S7). At the end of the reverse sweep, $\tilde{\mathbf{B}}^{(\sigma')}_{1}$ is updated, and the partial factorization vector is back to the form of Eq.(S2), which enables a regular sweep to be performed during the next run. Therefore, the fields are swept back and forth, and all partial factorizations not only save computational time but also provide direct access to unequal-time Green's functions or force terms at any imaginary time in Hybrid Monte Carlo algorithms.

For each update, one has to solve the eigenproblem $\mathbf{UDT} = \tilde{\mathbf{U}}\mathbf{D_C}\tilde{\mathbf{T}}$ as given in Eq.(S6)

$$\mathbf{UDTx} = \lambda \mathbf{x} \tag{S9}$$

to obtain the spectrum $\{\lambda\}$ that would be used in the canonical partition function calculation. A direct diagonalization of $\mathbf{UDT}$ is not numerically stable and one may permute the factorization as

$$\mathbf{DTUx'} = \lambda \mathbf{x'} \tag{S10}$$

with $\mathbf{x'} = \mathbf{U}^{-1}\mathbf{x}$. The resulting matrices $\mathbf{DTU}$ are row-graded matrices and can be stably diagonalized by matrix balancing [S5].

For the observable measurements presented in the main text, we initialize $10-60$ independent random walkers. Each walker first sweeps the lattice back and forth 100 times to thermalize and then collects approximately uncorrelated 1000 samples by measuring the observables every 20 sweep.

## II. PURITY AND FIDELITY CALCULATION DETAILS

For purity and fidelity measurements that require computing the ratio of partition functions as in Eqs.(??) and (??), we use the ensemble switching technique [S6, S7] that avoids direct evaluations of the partition functions. Without loss of generality, we define the expectation value of the target observable as a ratio

$$\langle r \rangle = \frac{Z}{Z'} = \frac{\int_\sigma Z(\sigma)}{\int_\sigma Z'(\sigma)}, \tag{S11}$$

where $\sigma$ denotes the HS field, and $Z$ and $Z'$ are two different abstract partition functions that can be either canonical or grand canonical and represent the partition functions of two different systems. The QMC estimate of Eq. (S11) can then be evaluated by first defining two transition probabilities between the different partition functions

$$\langle p_{Z' \to Z} \rangle = \frac{1}{Z'} \int_\sigma \min(1, \frac{Z(\sigma)}{Z'(\sigma)}) \cdot Z'(\sigma), \tag{S12}$$

$$\langle p_{Z \to Z'} \rangle = \frac{1}{Z} \int_\sigma \min(1, \frac{Z'(\sigma)}{Z(\sigma)}) \cdot Z(\sigma), \tag{S13}$$

and then taking the ratio of these two transition probabilities

$$\begin{aligned}\frac{\langle p_{Z' \to Z} \rangle}{\langle p_{Z \to Z'} \rangle} &= \frac{Z}{Z'} \frac{\int_\sigma \min(1, \frac{Z(\sigma)}{Z'(\sigma)}) \cdot Z'(\sigma)}{\int_\sigma \min(1, \frac{Z'(\sigma)}{Z(\sigma)}) \cdot Z(\sigma)} \\ &= \frac{Z}{Z'} \frac{\sum_{Z(\sigma) < Z'(\sigma)} Z(\sigma) + \sum_{Z(\sigma) > Z'(\sigma)} Z'(\sigma)}{\sum_{Z'(\sigma) < Z(\sigma)} Z'(\sigma) + \sum_{Z'(\sigma) > Z(\sigma)} Z(\sigma)} \\ &= \frac{Z}{Z'} = \langle r \rangle. \end{aligned} \tag{S14}$$

Hence, the measurement of the ratio of the partition functions is obtained by two separate QMC simulations using the same set of HS fields, with one sampling according to the weight $Z(\sigma)$ and measuring the transition probability $p_{Z \to Z'}$ and another sampling according to $Z'(\sigma)$ and measuring $p_{Z' \to Z}$. For purity calculations, this ensemble switching scheme turns out to have a faster convergence than the usual replica scheme [S6].

## III. FURTHER COMPARISONS BETWEEN THE RECURSION AND PROJECTION ALGORITHMS

### A. Comparison of Runtimes for the Calculation of Occupation Numbers in the Projection and Recursion Algorithms

Many common observables in many-body physics necessitate the computation of occupation numbers. In order to compute occupation numbers, the recursion and projection algorithms both need to compute the field-dependent one-body density in the diagonal space (occupation number), $\langle n_\alpha \rangle_N$, and then evaluate a one-body operator as $\langle \hat{c}_i^\dagger \hat{c}_j \rangle_N = \sum_\alpha \mathbf{P}_{i\alpha} \langle \hat{n}_\alpha \rangle_N \mathbf{P}^{-1}_{\alpha j}$, based on the eigendecomposition $\mathbf{U}_\sigma = \mathbf{P}\Lambda\mathbf{P}^{-1}$, where $\Lambda = \mathrm{diag}(\{\lambda_i\})$. In the projection formalism, the formula for the occupation number is thus given by

$$\langle n_\alpha \rangle_N = \frac{e^{-\beta \mu N}}{Z_N N_s} \sum_{m=1}^{N_s} e^{-i\phi_m N} \frac{e^{\beta\mu} e^{i\phi_m} \lambda_\alpha}{1 + e^{\beta\mu}e^{i\phi_m}\lambda_\alpha} \tilde{Z}_m. \tag{S15}$$

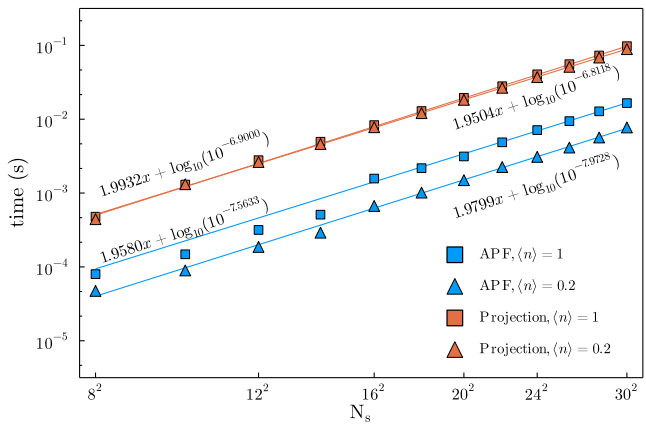

FIG. S1. Comparison of the runtime for computing the occupation number per sample of the Recursion algorithm against the Projection algorithm using the same dataset as the partition function benchmark in the main text.

The recursive equation for the occupation number is given in the main text.

Using these formulas, we can then compare the runtimes required by each of the algorithms to compute occupation numbers. As shown in Figure S1, computing occupation numbers using the projection algorithm is more expensive than doing so using the recursion algorithm by over one order of magnitude. Just as discussed in the main text for partition functions, the recursion algorithm becomes more efficient at lower particle fillings.

### B. Low-Temperature Speedups

At low temperatures, some values in $\{\lambda\}$ are extremely large or extremely small, and the corresponding $\langle n_\alpha \rangle_N$ values are close to 1 or 0, respectively. One can therefore neglect these nearly occupied or nearly unoccupied levels and only focus on the few active energy levels in the vicinity of the Fermi level. In the recursive formalism, this is achieved by truncating energy levels whose occupation probabilities are larger than a specified tolerance, $1 - \epsilon$,

$$\left| p_j^{(\mu)} \right| = \left| \frac{e^{\beta\mu}\lambda_j}{1 + e^{\beta\mu}\lambda_j} \right| > 1 - \epsilon, \quad (S16)$$

or whose "hole" occupation probabilities are smaller than $\epsilon$,

$$\left| 1 - p_j^{(\mu)} \right| = \left| \frac{1}{1 + e^{\beta\mu}\lambda_j} \right| < \epsilon. \quad (S17)$$

After the truncation, only $N_s^{(f)}$ active energy levels out of the total $N_s$ sites are left to perform the recursive calculation over the number of $N^{(f)} := N - N^{(o)}$ particles, where $N^{(o)}$ is the number of nearly occupied levels determined by Eq.(S17). The computational scaling is then reduced to $O(N_s^{(f)} N^{(f)})$.

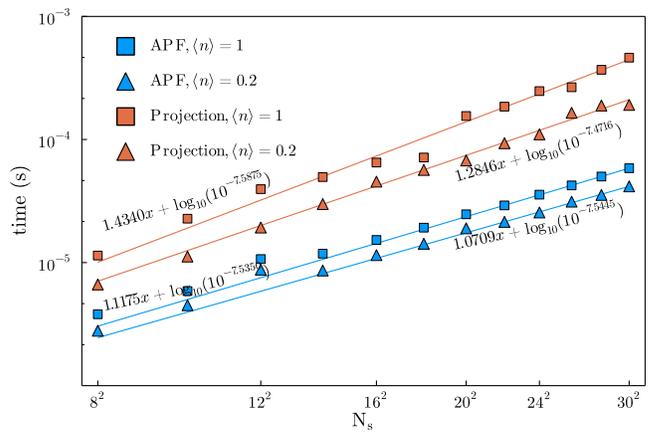

FIG. S2. Comparison of the runtime for computing the partition function per sample of the Recursion algorithm against the Projection algorithm using the low-temperature approximation. Number of energy levels that are involved in the calculation is chosen as the minimum value that satisfies the error tolerance $|\log Z_N^{\text{approx}} - \log Z_N| < 10^{-8}$

For the projection algorithm, the scaling is reduced by using fewer Fourier points in the Fourier sum[S8, S9], $N_{\text{FT}} < N_s$,

$$Z_N(N_{\text{FT}}) = \frac{1}{N_{\text{FT}}} \sum_{m=1}^{N_{\text{FT}}} e^{-\beta\mu N} e^{-i\phi_m N} \tilde{\mathcal{Z}}_\mu(m), \quad (S18)$$

with $\phi_m = 2\pi m / N_{\text{FT}}$. The error is given by

$$Z_N(N_{\text{FT}}) - Z_N = \frac{1}{e^{\beta\mu N}} \sum_{N', N'-N=kN_{\text{FT}}} e^{\beta\mu N'} Z_{N'}, \quad (S19)$$

which is based on the identity

$$\frac{1}{N_{\text{FT}}} \sum_{m=1}^{N_{\text{FT}}} e^{i\phi_m(N-N')} = \sum_{N', N'-N=kN_{\text{FT}}} \delta_{N,N'}. \quad (S20)$$

Note that the error depends on the value of $\beta$, therefore smaller $N_{\text{FT}}$ can be chosen for larger $\beta$ values, and the computational scaling is then reduced to $O(N_{\text{FT}}^2)$.

In Figure S2, the efficiencies of both algorithms after the approximation are tested using randomly-generated Hubbard AFQMC samples at $U = 2$ and $\beta = 20$, and the recursion algorithm is slightly more efficient than the projection algorithm as it roughly scales as $O(N_s)$. It is because, at low temperatures, the portion of "frozen" electrons is almost constant regardless of system size or particle filling, so the actual computational scaling is $O(N_s^{(f)})$, and hence the $\langle n \rangle = 0.2$ filling shows less improved speedup over the $\langle n \rangle = 1$ filling compared to the non-approximate benchmarks.

## C. Recursion and Projection Algorithms for Bosons

We derive a particle-number projection formalism for the bosonic canonical partition function in the AFQMC framework via the Fourier transform. This is accomplished with the projection operator

$$\hat{P}_N = \frac{1}{2\pi} \int_0^{2\pi} d\phi e^{i\phi(\hat{N}-N)}. \quad \text{(S21)}$$

Similar to the fermionic AFQMC formalism, if we define the field-dependent bosonic propagator as $\hat{B}^{(\varphi)}$, then the bosonic canonical partition function for a specified auxiliary-field $\varphi$ reads

$$Z_N(\varphi) = \text{Tr}[\hat{P}_N \hat{B}] = \frac{1}{2\pi}\int_0^{2\pi} d\phi e^{-i\phi N}\text{Tr}[e^{i\phi\hat{N}}\hat{B}^{(\varphi)}] \quad \text{(S22a)}$$

$$= \frac{1}{2\pi}\int_0^{2\pi} d\phi e^{-i\phi N}\det[(\mathbf{I} - e^{i\phi}\mathbf{B}^{(\varphi)})^{-1}] \quad \text{(S22b)}$$

$$= \frac{1}{2\pi}\int_0^{2\pi} d\phi \frac{e^{-i\phi N}}{\det[\mathbf{I} - e^{i\phi}\mathbf{B}^{(\varphi)}]}, \quad \text{(S22c)}$$

where we have used the identity

$$\text{Tr}[e^{\hat{X}}e^{\hat{Y}}] = \det[(\mathbf{I} - e^{\mathbf{X}}e^{\mathbf{Y}})^{-1}], \quad \text{(S23)}$$

with $\hat{X}, \hat{Y}$ being one-body operators for bosons.

Note that the numerical evaluation of Eq.(S22c) can be simplified by computing the eigenvalues, $\lambda_i$, of the matrix $\mathbf{B}^{(\varphi)}$. Using these eigenvalues, Eq.(S22c) can be written as

$$Z_N(\varphi) = \frac{1}{2\pi}\int_0^{2\pi} d\phi \frac{e^{-i\phi N}}{\prod_i^{N_s}(1 - e^{i\phi}\lambda_i)}. \quad \text{(S24)}$$

To prevent large values of $\lambda_i$ from destabilizing the calculation, we can rescale them by a factor $e^{\beta\mu}$ (i.e. $\lambda_i' = e^{\beta\mu}\lambda_i$), as was done in the fermionic case, and the equation becomes

$$Z_N(\varphi) = \frac{1}{2\pi}\int_0^{2\pi} d\phi \frac{e^{-i\phi N}e^{-\beta\mu N}}{\prod_i^{N_s}(1 - e^{i\phi}e^{\beta\mu}\lambda_i)}. \quad \text{(S25)}$$

One can then approximate the integral by discretization

$$Z_N(\varphi) \approx \frac{1}{M}\sum_{i=0}^{M} \frac{e^{-i\phi_i N}e^{-\beta\mu N}}{\prod_i^{N_s}(1 - e^{i\phi_i}e^{\beta\mu}\lambda_i)}, \quad \text{(S26)}$$

with $\phi_i = 2\pi i/M$. Obviously the number of discretization bins should be larger than the particle number (i.e. $M > N$), and numerical experiments show that the series converge for values of $M > 10N$.

The projection method for bosons has several drawbacks compared to the recursive method. First, the computational scaling is $O(MN)$ for the Projection method, while for the Recursive method, a direct calculation of the partition function can be avoided by using the relations [S10],

$$\begin{cases} \langle n_\alpha \rangle_{N+1} = \frac{Z_N}{Z_{N+1}}\lambda_\alpha(1 + \langle n_\alpha \rangle_N), \\ \frac{Z_N}{Z_{N+1}} = \frac{N+1}{\sum_\alpha \lambda_\alpha(1+\langle n_\alpha \rangle_N)}, \end{cases} \quad \text{(S27)}$$

and the numerical effort to compute the partition function and the occupation numbers only scales as $O(N)$. Unlike the fermionic case, the Bosonic recursion is highly stable and no rescaling factor (chemical potential) as in Eq.(S26) is needed to stabilize the calculation. Since the chemical potential in Eq.(S26) is not as easy to specify for bosons as it is for fermions, one has to solve the equation

$$\frac{\text{Tr}[e^{\beta\mu}\hat{B}\hat{N}]}{\text{Tr}[e^{\beta\mu}\hat{B}]} = \sum_i \frac{1}{(e^{\beta\mu}\lambda_i)^{-1} - 1} = N \quad \text{(S28)}$$

to pinpoint $e^{\beta\mu}$, and the numerical solution may not be stable for large particle numbers. Recall that the Bose-Einstein distribution

$$\langle n_i \rangle_\mu = \frac{1}{e^{\beta(\epsilon_i-\mu)} - 1} = \frac{1}{(e^{\beta\mu}\lambda_i)^{-1} - 1} \quad \text{(S29)}$$

requires that the chemical potential does not exceed the ground state energy $\epsilon_0$. For cases where the particle number is sufficiently large such that $e^{\beta\mu}\lambda_i$ is quite close to 1, the Projection algorithm becomes numerically unstable.

## IV. DIFFERENCES IN STATE OCCUPATIONS BETWEEN ENSEMBLES

In the main text, we examined differences in the charge structure factor between the two different ensembles. Here, we similarly examine the differences between occupation numbers in the two different ensembles. Fig. S3 illustrates the average occupation number at different momenta for a 2D Hubbard model at filling $\langle n \rangle = \frac{2 \times 23}{36}$ with $U = 2$. Deviations are visible in the occupation numbers between the canonical and grand canonical ensembles close to the Fermi energy. As in the simple case of only one particle in two levels, demanding that occupations be equal would require shifting the grand canonical occupations to a larger value of $\beta$.

This behavior can be quantified by integrating deviations between the occupation numbers over all momenta, which for a discrete spectrum is given by

$$\chi(\beta_N, \beta_\mu) = \sum_{\boldsymbol{k}} \left|\langle n_{\boldsymbol{k}}\rangle_N - \langle n_{\boldsymbol{k}}\rangle_\mu\right|, \quad \text{(S30)}$$

where each expectation value could have its own temperature to minimize deviations (as above) and $\mu$ has been fine-tuned such that the average number of particles in the grand canonical ensemble is equal to $N$. The results

4

for the 6 × 6 Hubbard model at $U = 2$ away from half-filling ($\langle n \rangle = 1.28$) are shown in a logarithmically-scaled heat-map in Fig. S4.

Complete agreement between ensembles would be signalled by a minimum in $\chi$ at $\beta_\mu = \beta_N$ (dashed line), but instead we observe considerable deviations that persist and grow at large values of $\beta$. While the magnitude of the color scale depends on the system size, one can always find a temperature region for any finite size system where high precision thermometry protocols necessitate the use of the correct ensemble.

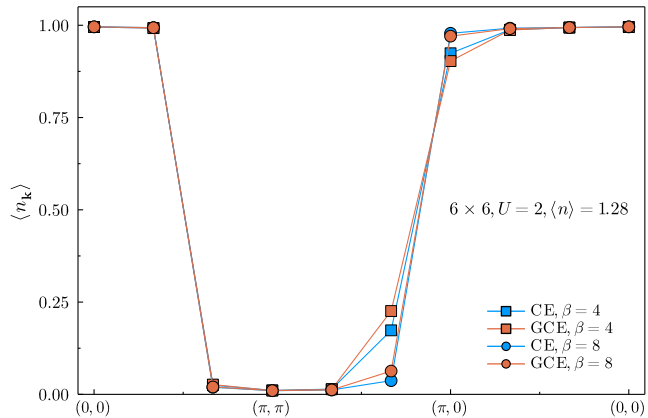

FIG. S3. Average momentum distribution as a function of $\vec{k}$ in the canonical and grand canonical ensembles for a 6×6 Hubbard model at $U = 2$, $\langle n \rangle = 1.28$, and either $\beta = 4$ or $\beta = 8$.

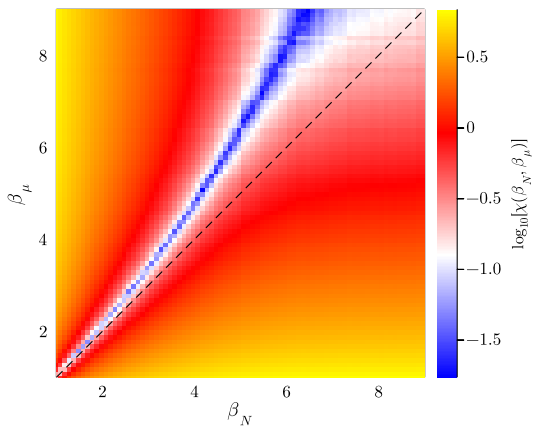

FIG. S4. Heatmap of $\log_{10}[\chi(\beta_N, \beta_\mu)]$, the difference between the occupation of different momentum modes as a function of the temperature in the canonical and grand canonical ensembles. The dashed line denotes the diagonal at which $\beta_\mu = \beta_N$. The upward-sloping blue curve indicates that lower temperature (larger $\beta_\mu$) GCE momentum distributions agree most with higher temperature (smaller $\beta_N$) CE momentum distributions.

## V. PURITY AND FIDELITY OF THERMAL STATES

Here, we compare the purity of the thermal state in the canonical and the grand canonical ensembles. To do so, first, we review the relationship between the canonical and the grand canonical thermal states in a system of itinerant particles. If the Hamiltonian $\hat{H}$ governs a system of itinerant particles and also commutes with the system number operator $\hat{N}$, then, we can write

$$e^{-\beta \hat{H}} = \sum_N \hat{\Pi}_N e^{-\beta \hat{H}} \hat{\Pi}_N, \quad (S31)$$

and thus

$$\text{Tr}\left[e^{-\beta \hat{H}}\right] = \sum_N \text{Tr}_N\left[e^{-\beta \hat{H}}\right], \quad (S32)$$

where, in general, the canonical trace $\text{Tr}_N$ is defined as $\text{Tr}_N[\hat{\rho}] = \text{Tr}\left[\hat{\Pi}_N \hat{\rho} \hat{\Pi}_N\right]$. Accordingly, the canonical thermal state is

$$\hat{\rho}_N(\beta) = \frac{\hat{\Pi}_N e^{-\beta \hat{H}} \hat{\Pi}_N}{Z_N(\beta)}, \quad (S33)$$

where $Z_N(\beta) = \text{Tr}_N\left[e^{-\beta \hat{H}}\right]$ is the canonical partition function. Next, the thermal state of the system in the grand canonical ensemble is

$$\hat{\rho}_\mu(\beta) = \frac{e^{-\beta(\hat{H} - \mu \hat{N})}}{\mathcal{Z}_\mu(\beta)}, \quad (S34)$$

where, $\mathcal{Z}_\mu(\beta) = \text{Tr}\left[e^{-\beta(\hat{H} - \mu \hat{N})}\right]$ is the grand canonical partition function and the constant $\mu$ is the chemical potential. As a generalization of Eq. (S31), we write

$$e^{-\beta(\hat{H} - \mu \hat{N})} = \sum_N \hat{\Pi}_N e^{-\beta \hat{H}} \hat{\Pi}_N e^{\beta \mu N}, \quad (S35)$$

and thus,

$$\hat{\rho}_\mu(\beta) = \sum_N P_\mu(N, \beta) \hat{\rho}_N(\beta), \quad (S36)$$

where

$$P_\mu(N; \beta) = \frac{e^{\beta \mu N} Z_N(\beta)}{\mathcal{Z}_\mu(\beta)}, \quad (S37)$$

is the probability that the system is in the $N$-particle sector for a given $\mu$ and $\beta$.

### A. Purity

The purity of a normalized quantum state $\hat{\rho}$ is defined as

$$\mathcal{P} = \text{Tr}\left[\hat{\rho}^2\right]. \quad (S38)$$

In general, $\mathcal{P}$ is bounded as $1/|\mathcal{H}| \leq \mathcal{P} \leq 1$, where $|\mathcal{H}|$ is the dimension of the corresponding Hilbert space $\mathcal{H}$. The upper bound can be reached, only if $\hat{\rho}$ is pure, in which case it can be expressed as $\hat{\rho} = |\Psi\rangle\langle\Psi|$, where $|\Psi\rangle$ is a normalized state. The lower bound is achieved if $\hat{\rho}$ is proportional to the identity $\hat{I}$, i.e., $\hat{\rho} = \frac{1}{|\mathcal{H}|}\hat{I}$. Consider the thermal state $\hat{\rho}(\beta) = \frac{e^{-\beta\hat{H}}}{\text{Tr}[e^{-\beta\hat{H}}]}$ of a system at an inverse temperature $\beta$, where $\hat{H}$ is the system Hamiltonian. The purity of such a state is given by

$$\mathcal{P} = \frac{Z(2\beta)}{Z(\beta)^2}, \tag{S39}$$

where $Z(\beta) = \text{Tr}\left[e^{-\beta\hat{H}}\right]$ is the corresponding partition function.

In the canonical ensemble, we have

$$\mathcal{P}_N = \frac{Z_N(2\beta)}{Z_N(\beta)^2}, \tag{S40}$$

or equivalently

$$\ln \mathcal{P}_N = 2\beta \left[F_N(\beta) - F_N(2\beta)\right], \tag{S41}$$

where $F_N(\beta) = -\ln Z_N(\beta)/\beta$ is the free energy. However, the purity of the grand canonical thermal state is

$$\mathcal{P}_\mu = \frac{\mathcal{Z}_\mu(2\beta)}{\mathcal{Z}_\mu(\beta)^2}, \tag{S42}$$

and similarly

$$\ln \mathcal{P}_\mu = 2\beta \left[\Phi_\mu(\beta) - \Phi_\mu(2\beta)\right], \tag{S43}$$

where $\Phi_\mu(\beta) = -\ln \mathcal{Z}_\mu(\beta)/\beta$ is the grand free energy.

We can relate the canonical and the grand canonical purity using Eq (S37), where

$$\mathcal{P}_\mu = \mathcal{P}_N \frac{P_\mu(N;\beta)^2}{P_\mu(N;2\beta)}. \tag{S44}$$

### B. Fidelity

Given a quantum state $\hat{\rho}$, we can measure how close it is to a target state $\hat{\rho}_t$ using the mixed-state quantum fidelity $\mathcal{F}(\hat{\rho}, \hat{\rho}_t)$, where $\mathcal{F}$ measures the similarity between the two quantum states. Yet, in the context of quantum information, different measures of mixed-state fidelity have been proposed [S11–S14]. The most familiar definition of the mixed-state fidelity is [S11, S12]

$$\tilde{\mathcal{F}}(\hat{\rho}, \hat{\rho}_t) = \text{Tr}\left[\sqrt{\sqrt{\hat{\rho}_t}\hat{\rho}\sqrt{\hat{\rho}_t}}\right], \tag{S45}$$

which, in general, is difficult to calculate. However a much simpler expression has been proposed [S13]

$$\mathcal{F}(\hat{\rho}, \hat{\rho}_t) = \frac{|\text{Tr}[\hat{\rho}\hat{\rho}_t]|}{\sqrt{\text{Tr}[\hat{\rho}^2]\text{Tr}[\hat{\rho}_t^2]}}. \tag{S46}$$

To test how the grand canonical states approximate the canonical states, we compute

$$\mathcal{F}(\hat{\rho}_\mu, \hat{\rho}_N) = \frac{\text{Tr}[\hat{\rho}_\mu \hat{\rho}_N]}{\sqrt{\text{Tr}[\hat{\rho}_\mu^2]\text{Tr}[\hat{\rho}_N^2]}}, \tag{S47}$$

which, using $\hat{\rho}_\mu \hat{\rho}_N = P_\mu(N;\beta)\hat{\rho}_N$, gives

$$\mathcal{F}(\hat{\rho}_\mu, \hat{\rho}_N) = P_\mu(N;\beta)\sqrt{\frac{\text{Tr}[\hat{\rho}_N^2]}{\text{Tr}[\hat{\rho}_\mu^2]}} = P_\mu(N;\beta)\sqrt{\frac{\mathcal{P}_N}{\mathcal{P}_\mu}}. \tag{S48}$$

Furthermore, using Eq (S44), we get

$$\mathcal{F}(\hat{\rho}_\mu, \hat{\rho}_N) = \sqrt{P_\mu(N;2\beta)}. \tag{S49}$$

Also, employing the fact that $[\hat{\rho}_\mu, \hat{\rho}_N] = 0$, simplifies $\tilde{\mathcal{F}}(\hat{\rho}_\mu, \hat{\rho}_N)$ to

$$\tilde{\mathcal{F}}(\hat{\rho}_\mu, \hat{\rho}_N) = \sqrt{P_\mu(N;\beta)}. \tag{S50}$$

## VI. LOW TEMPERATURE BEHAVIOR OF THE PURITY AND FIDELITY

Let the the spectrum of $\hat{H}$ be $\{E_{N,i}\}$ and the corresponding degeneracies be $\{g_{N,i}\}$, where $\{E_{N,i}\} > \{E_{N,j}\}$ for $i > j$. The canonical partition function $Z_N(\beta)$ for $N$ particles at some inverse temperature $\beta$ is then

$$Z_N(\beta) = \sum_i g_{N,i} e^{-\beta E_{N,i}}. \tag{S51}$$

For $\beta\Delta_{N,1} \gg 1$, where $\Delta_{N,i} = E_{N,i} - E_{N,0}$, we have $Z_N(\beta) \approx g_{N,0} e^{-\beta E_{N,0}} \left(1 + \frac{g_{N,1}}{g_{N,0}} e^{-\beta\Delta_{N,1}} + \mathcal{O}(e^{-\beta\Delta_{N,2}})\right)$, and thus

$$\mathcal{P}_N = \frac{Z_N(2\beta)}{Z_N(\beta)^2}$$
$$\approx \frac{g_{N,0} e^{-2\beta E_{N,0}}\left(1 + \frac{g_{N,1}}{g_{N,0}} e^{-2\beta\Delta_{N,1}} + \mathcal{O}(e^{-2\beta\Delta_{N,2}})\right)}{g_{N,0}^2 e^{-2\beta E_{N,0}}\left(1 + \frac{g_{N,1}}{g_{N,0}} e^{-\beta\Delta_{N,1}} + \mathcal{O}(e^{-\beta\Delta_{N,2}})\right)^2}$$
$$\approx \frac{1}{g_{N,0}}\left(1 - \frac{2g_{N,1}}{g_{N,0}} e^{-\beta\Delta_{N,1}} + \mathcal{O}(e^{-\beta\tilde{\Delta}_N})\right), \tag{S52}$$

where $\tilde{\Delta} = \min\{\Delta_{N,2}, 2\Delta_{N,1}\}$. For a non-degenerate ground state, $g_{N,0} = 1$, we have

$$\ln(1 - \mathcal{P}_N) \approx -\beta\Delta_{N,1} + \ln 2g_{N,1}. \tag{S53}$$

For the grand canonical case,

$$\mathcal{Z}_\mu(\beta) = \sum_N e^{\beta\mu N} Z_N(\beta), \tag{S54}$$

where at sufficiently large $\beta$, we have $\mathcal{Z}_\mu(\beta) \approx e^{\beta\mu N}\left(e^{-\mu\beta}Z_{N-1}(\beta) + Z_N(\beta) + e^{\mu\beta}Z_{N+1}(\beta)\right)$ (probability distribution argument) or

$$\mathcal{Z}_\mu(\beta) \approx e^{\beta\mu N}Z_N(\beta)\left(1 + e^{-\mu\beta}\frac{Z_{N-1}(\beta)}{Z_N(\beta)} + e^{\mu\beta}\frac{Z_{N+1}(\beta)}{Z_N(\beta)}\right). \quad (S55)$$

Also,

$$\frac{Z_{N+1}(\beta)}{Z_N(\beta)} \approx \frac{g_{N+1,0}}{g_{N,0}}e^{-\beta(E_{N+1,0}-E_{N,0})}\frac{1+\mathcal{O}(e^{-\beta\Delta_{N+1,1}})}{1+\mathcal{O}(e^{-\beta\Delta_{N,1}})}$$
$$\approx \frac{g_{N+1,0}}{g_{N,0}}e^{-\beta(E_{N+1,0}-E_{N,0})}, \quad (S56)$$

and similarly

$$\frac{Z_{N-1}(\beta)}{Z_N(\beta)} \approx \frac{g_{N-1,0}}{g_{N,0}}e^{-\beta(E_{N-1,0}-E_{N,0})}. \quad (S57)$$

Putting everything together, we get

$$\mathcal{Z}_\mu(\beta) \approx e^{\beta\mu N}Z_N(\beta)\left[1 + \frac{g_{N-1,0}}{g_{N,0}}e^{\beta(E_{N,0}-E_{N-1,0}-\mu)}\right.$$
$$\left.+ \frac{g_{N+1,0}}{g_{N,0}}e^{-\beta(E_{N+1,0}-E_{N,0}-\mu)}\right]. \quad (S58)$$

Substituting this into Eq. (S42) yields

$$\mathcal{P}_\mu \approx \mathcal{P}_N[1 - 2\frac{g_{N-1,0}}{g_{N,0}}e^{\beta(E_{N,0}-E_{N-1,0}-\mu)}$$
$$- 2\frac{g_{N+1,0}}{g_{N,0}}e^{-\beta(E_{N+1,0}-E_{N,0}-\mu)}]. \quad (S59)$$

### A. Fidelity

From Eq.(S58), we can approximate $P_\mu(N;\beta)$ as (assuming that the probability distribution is sharply peaked at an average of $N$)

$$P_\mu(N;\beta) = \frac{e^{\beta\mu N}Z_N(\beta)}{\mathcal{Z}_\mu(\beta)}$$
$$\approx [1 + \frac{g_{N-1,0}}{g_{N,0}}e^{\beta(E_{N,0}-E_{N-1,0}-\mu)}$$
$$+ \frac{g_{N+1,0}}{g_{N,0}}e^{-\beta(E_{N+1,0}-E_{N,0}-\mu)}]^{-1}$$
$$\approx 1 - \frac{g_{N-1,0}}{g_{N,0}}e^{\beta(E_{N,0}-E_{N-1,0}-\mu)}$$
$$- \frac{g_{N+1,0}}{g_{N,0}}e^{-\beta(E_{N+1,0}-E_{N,0}-\mu)}, \quad (S60)$$

and thus

$$\mathcal{F}(\hat{\rho}_\mu, \hat{\rho}_N) = \sqrt{P_\mu(N;2\beta)}$$
$$\approx [1 - \frac{g_{N-1,0}}{g_{N,0}}e^{2\beta(E_{N,0}-E_{N-1,0}-\mu)}$$
$$- \frac{g_{N+1,0}}{g_{N,0}}e^{-2\beta(E_{N+1,0}-E_{N,0}-\mu)}]^{1/2}$$
$$\approx 1 - \frac{g_{N-1,0}}{2g_{N,0}}e^{2\beta(E_{N,0}-E_{N-1,0}-\mu)}$$
$$- \frac{g_{N+1,0}}{2g_{N,0}}e^{-2\beta(E_{N+1,0}-E_{N,0}-\mu)}. \quad (S61)$$

Combining the two exponential terms then gives

$$\ln[1 - \mathcal{F}(\hat{\rho}_\mu, \hat{\rho}_N)] \approx -2\beta\Delta\tilde{E} + 2\ln\frac{g}{2}, \quad (S62)$$

with $\Delta\tilde{E}$ and $g$ represent the effective gap and degeneracy, respectively.


\fi

\end{document}